\documentclass[11pt,a4paper]{article}
\usepackage[colorlinks]{hyperref}
\usepackage[utf8]{inputenc}
\usepackage[english]{babel}
\usepackage{amsmath}
\usepackage{amsfonts}
\usepackage{amssymb}
\usepackage{bm}
\usepackage{graphicx}
\usepackage{lmodern}
\usepackage{tabulary}
\usepackage{multirow}
\usepackage{color}
\usepackage[margin=2.5cm]{geometry}
\usepackage[affil-it]{authblk}
\newtheorem{theorem}{Theorem}[section]
\newtheorem{proposition}{Proposition}[section]

\title{Linear models for the impact of order flow on prices \\ II. The Mixture Transition Distribution model}
\author[1]{Damian Eduardo Taranto}
\author[2]{Giacomo Bormetti}
\author[3]{Jean-Philippe Bouchaud}
\author[1]{\\Fabrizio Lillo}
\author[3]{Bence T\'{o}th}

\affil[1]{Scuola Normale Superiore, Piazza dei Cavalieri 7, 56126 Pisa, Italy}
\affil[2]{Department of Mathematics, University of Bologna, Piazza di Porta San Donato 5, 40126 Bologna, Italy}
\affil[3]{Capital Fund Management, 23-25, Rue de l'Universit\'e 75007 Paris, France}
\vspace{0.5cm}
\begin{document}
\maketitle
\begin{abstract}
  \noindent Modeling the impact of the order flow on asset prices is of primary importance to understand the behavior of financial markets. Part I of this paper reported the remarkable improvements in the description of the price dynamics which can be obtained when one incorporates the impact of past returns on the future order flow. However, impact models presented in Part I consider the order flow as an exogenous process, only characterized by its two-point correlations. This assumption seriously limits the forecasting ability of the model. Here we attempt to model directly the stream of discrete events with a so-called 
  Mixture Transition Distribution (MTD) framework, introduced originally by Raftery (1985). We distinguish between price-changing and non price-changing events and combine them with the order sign in order to reduce the order flow dynamics to the dynamics of a four-state discrete random variable. The MTD represents a parsimonious approximation of a full high-order Markov chain. The new approach captures with adequate realism the conditional correlation functions between signed events for both small and large tick stocks and signature plots. From a methodological viewpoint, we discuss a novel and flexible way to calibrate a large class of MTD models with a very large number of parameters. In spite of this large number of parameters, an out-of-sample analysis confirms that the model 
  does not overfit the data. 
\end{abstract}

\section{Introduction}
This paper is the companion of~\cite{taranto16}. In the previous, part I paper we discussed the differences and similarities between two linear models describing the impact of order flow on prices, namely the Transient Impact Model (TIM) and the History Dependent Impact Model (HDIM). In these models, the sign of the order flow is considered to be an exogenous, time correlated process that affects price dynamics either through a ``propagator'', i.e. a linear combination of past values (TIM) or via a ``surprise'' mechanism, i.e. the deviation between the realised order flow and its expected level (HDIM). In reality, however, order flow is not exogenous and is itself affected by the past history of price. In~\cite{taranto16} we partly overcame this issue by enhancing the description of the order flow to account for price changing events and non price changing events, in the spirit of~\cite{eisler2012a,eisler2012b}. This allows one to encode the propensity of the order flow to invert its sign after a price change, an effect that is particularly important for large tick stocks. This extended model improves significantly the description of the price process, both in terms of the lag-dependent volatility (i.e. the signature plot) and in terms of the response function computed for negative lags. However this approach is incomplete as it does not specify the data generating process for the order flow itself, which is only described through two-point correlation functions. This does not allow one to {\it forecast} the future order flow itself, for example whether a trade is likely to change the price or not. 

In this paper we attempt to model the joint dynamics of order flow and prices. This family of models has a long tradition in market microstructure, starting from the seminal work of Hasbrouck~\cite{hasbrouck88,hasbrouck91}, who proposed a Vector Autoregressive (VAR) model for the joint dynamics of order flow and prices\footnote{More recent modeling in continuous time makes use of Hawkes processes~\cite{bacry14}, which bear some degree of similarity with the models considered in the present paper.}. There are two main related limitations of this approach. The first is that VAR models are adequate for variables with continuous support (e.g. Gaussian), while the order flow (signs and events) and tick by tick price changes are more naturally described by discrete variables. Second, the standard VAR approach prescribes a linear relation between the variables, while a broader definition includes the possibility of a linear relation between past variables and the {\it probability} of observing in the future the value of a given variable. 

A paradigmatic example is a finite state Markov chain $X_t$. Let $m$ be the number of states, $\boldsymbol{Q}$ the $m\times m$ time-invariant transition matrix, and let $\chi_t=(x_t(1),...,x_t(m))$ be a row vector such that $x_t(i)=1$ if $X_t=i$ and zero otherwise. The probability vector $\hat\chi_t=(\mathbb{P}(X_t=1),..., \mathbb{P}(X_t=m))$ is determined by the {\it linear} system of equations
\begin{equation*}
  \hat\chi_t= \chi_{t-1} \boldsymbol{Q}\,.
\end{equation*}
Therefore a natural way to describe the joint dynamics of discrete valued variables (such as the order flow sign and price changes) in a linear setting is with a Markov process of large order. In fact, we have shown in Ref.~\cite{taranto16} that for large tick stocks\footnote{We remind that large tick stocks have the property that the ratio between tick size and price is relatively high and as a consequence spread is almost always equal to one tick.} the model with two propagators (TIM2) corresponding to price changing and not-changing trades gives constant (in time) propagators when calibrated on real data (see top left panel of Fig.~7 of  ~\cite{taranto16}). This means that the knowledge of the order flow and the information on whether a trade changes the price completely characterises the price dynamics.  Thus, in the framework of linear models, it is natural to describe the system with a Markov process with $m=4$ states, $(\epsilon_t,\pi_t)\in\lbrace(-1,\mathrm{C}),(-1,\mathrm{NC}),(+1,\mathrm{NC}),(+1,\mathrm{C})\rbrace$, corresponding to buys ($\epsilon_t=+1$) and sells ($\epsilon_t=-1$) and price changing ($\pi_t=\mathrm{C}$) and not changing ($\pi_t= \mathrm{NC}$) trades. 

However, the main limitation of Markov models comes from the long memory of the order flow~\cite{bouchaud2004,lillo2004}. Since the order flow sign is very persistent, a low order Markov process  cannot be suitable to describe real markets. On the other hand, Markov processes of high order $p$ depend in general on a very large number of parameters ($O(m^p)$) and might result in inefficient estimation when a limited amount of data is available. For this reason in this paper we propose to use a parsimonious, yet versatile class of high order Markov processes termed the  Mixed Transition Distribution (MTD) model~\cite{raftery1985}  and its generalization (MTDg)~\cite{berchtold1995}. Thanks to a simple structure, where each lag contributes to the prediction of the current state in a separate and additive way, the dimension of model parameter space grows only linearly with the order of the MTDg model, i.e. as $O(m^2p)$. The model can be calibrated via Maximum Likelihood or via the Generalized Method of Moments. Moreover in the case of $m=2$ states (such as the signs of the order flow), the version of the MTDg model proposed in this paper reduces to the Discrete Autoregressive (DAR) model~\cite{jacobs1978}, which has been used to model the order flow in~\cite{taranto2014} and in the companion of this paper~\cite{taranto16}. Hence MTD and MTDg aim at providing a natural generalisation of the DAR(p) model to account for an arbitrary number of $m \geq 2$ states, while avoiding the exponential increase ($m^p$) of the number of parameters of the full Markov model. Perhaps surprisingly, this class of models has not been applied to financial data and the present paper attempts to fill this gap. In fact, the main methodological innovation of our work is a parametrization of the MTDg model which can be estimated even when the number of parameters is very large, as required to account for the correlation structure of financial data.

Specifically, we consider in this paper MTD and MTDg models as promising models for the joint dynamics of order flow and price changes for large tick stocks. Compared to the models investigated in Ref.~\cite{taranto16}, we provide here an explicit model for the order flow, and in particular its response to past price dynamics. Thus we aim at reproducing with the MTD model the complex conditional correlation functions of signed events for large tick stocks (see left panel of Fig.~4 of~\cite{taranto16} whose curves are reproduced also in Fig.~\ref{fig:sim_mtd_exp_eps_msft}). Moreover our modeling approach allows to perform out of sample analyses of the MTD's forecasting ability of the order flow and future price changes. Still, this framework has limitations when calibrated on anonymized order flow because one cannot easily disentangle order flow correlations coming from ``herding'' and coming from ``order splitting''. In other words, although MTDs give explicit predictions for the response of the order flow to a single event (impulse response), one has to be careful in interpreting the result, as it might not describe the true reaction of the market to an isolated, exogeneous order (see~\cite{toth2015,toth2012, toth2016}). 


The paper is organized as follows. In Section \ref{sec:mtd} we review the definition, main properties, and estimation methods of MTD and MTDg. In Section \ref{sec:mtdmicro} we present our parametrization of the model and some proposed improvements for the estimation. This Section also contains our empirical results on real financial data. Section \ref{sec:out} describes the results of some out of sample analyses for predicting price changes and order flow and in Section \ref{sec:conclusions} we draw some remarks and conclusions, and discuss the limitations of our approach.

\section{The Mixture Transition Distribution model}
\label{sec:mtd}

\subsection{Definition}
We start from a simple, but restrictive, definition of MTD models. Let $\left\lbrace X_t\right\rbrace_{t\in \mathbb{N}}$ be a sequence of random variables taking values in the finite set $\mathcal{X}=\lbrace 1,\ldots,m\rbrace$. This random sequence is said to be a $p$-th order MTDg sequence if for all $t > p$ and for all $(i,i_1,\ldots,i_p) \in \mathcal{X}^{p+1}$,
\begin{align}
  \mathbb{P}(X_t=i|X_{t-1}=i_1,\ldots,X_{t-p}=i_p) = \sum_{g=1}^p \lambda_g q^g_{i_g,i}\,,
  \label{eqn:mtdg}
\end{align}
where the vector $\lambda=(\lambda_1,\ldots,\lambda_p)$ is subject to the constraints:
\begin{align}
  \lambda_g &\geq 0, \qquad \forall g \in \lbrace 1,\ldots,p\rbrace\,, \label{eqn:mtd_cond1} \\
  \sum_{g=1}^p \lambda_g&=1\,. \label{eqn:mtd_cond2}
\end{align}
The matrices $\left\lbrace \boldsymbol{Q}^g=\left[q_{i,j}^g\right]; \, {i,j\in \mathcal{X}}; 1 \leq g \leq p \right\rbrace$ are positive $m \times m$ stochastic matrices, i.e. they satisfy
\begin{align}
  q_{i,j}^g \ge 0 \quad \mbox{and} \quad \sum_{j=1}^m q_{i,j}^g=1 \qquad \forall g \in \lbrace 1,\ldots,p\rbrace, \forall i,j\in \mathcal{X}\,. \label{eqn:mtd_cond3}
\end{align}

Raftery~\cite{raftery1985} has originally defined the model with the same transition matrix $\boldsymbol{Q}^g\equiv\boldsymbol{Q}$ for each lag $g=1,\ldots,p$ and this model is called the MTD. Later, Berchtold~\cite{berchtold1995} has introduced the more general definition of MTD models as a mixture of transitions from subsets of lagged variables $\lbrace X_{t-1},\ldots,X_{t-p} \rbrace$ to the present one $X_t$. In other words, the order of the transition matrices $\boldsymbol{Q}^g$ can be larger than one. 

Berchtold and Raftery~\cite{berchtold2002} have published a complete review of the MTD model. They recall theoretical results on the limiting behavior of the model and on its auto-correlation structure. In particular, they proved that if conditions of Eqs.~\ref{eqn:mtd_cond1}, \ref{eqn:mtd_cond2}, and \ref{eqn:mtd_cond3} are satisfied, then the model of Eq.~\ref{eqn:mtdg} is a well defined high-order Markov chain and its stationary distribution $\hat{\eta}=(\hat{\eta}_1,\ldots,\hat{\eta}_m)$ exists and it is unique. The above Mixture Transition Distribution models are Markov models where each lag $X_{t-1}, X_{t-2}, \ldots$ contributes additively to the distribution of the random variable $X_t$. Hence the model is linear in the sense described in the introduction.

In words, this class of models means the following: in order to determine the type of event $X_t$, occurring at time $t$, start choosing a reference time $t-g$ in the past, where $g$ is drawn at random with probability $\lambda_g$. If the event $X_{t-g}$ that occurred at time $t-g$ is of type $j$, then choose the event at time $t$ to be of type $i$ with probability $q_{i,j}$. This model can thus be interpreted as a probabilistic mixture of Markov processes. For this interpretation the fact that $(\lambda_g)_{g=1,...,p}$ is a probability vector and $\boldsymbol{Q}^g$ are stochastic matrices is critical. However, as already noted in the original papers~\cite{raftery1985,berchtold1995}, the MTDg model can be also defined when these parameters are negative or larger than one, provided that the conditions 
\begin{align}
0 \le \sum_{g=1}^p \lambda_g q_{i_g,i}^g \le 1, \qquad \forall (i,i_1,\ldots i_p)\in \mathcal{X}^{p+1}\,,
\end{align}
are satisfied, in such a way that all transition probabilities are well defined. As we shall see below, calibrated parameters do not necessarily abide to the
probabilistic interpretation. 

In this paper we will consider a specific class of MTDg models where the matrices $\boldsymbol{Q}^g$ share the same stationary state, i.e. the same left eigenvector $\hat{\eta}$ corresponding to the eigenvalue $1$. Under this assumption, generalizing a result of~\cite{berchtold1995}, we can prove the following theorem of the existence and uniqueness of the stationary distribution.
\begin{theorem}\label{teo1}
Suppose that a sequence of random variables $\left\lbrace X_t\right\rbrace_{t\in \mathbb{N}}$ taking values in the finite set $\mathcal{X}=\lbrace 1,\ldots,m\rbrace$ is defined by 
\begin{align*}
\mathbb{P}(X_t=i|X_{t-1}=i_1,\ldots,X_{t-p}=i_p) = \sum_{g=1}^p \lambda_g q^g_{i_g,i}\,,
\end{align*}
where $\boldsymbol{Q}^g=\left[q_{i,j}^g\right]_{i,j\in \mathcal{X}}$ are matrices with normalized rows, $\sum_j q^g_{i,j}=1$,
$\sum_{g=1}^p \lambda_g=1$, and assume that $\hat{\eta} \boldsymbol{Q}^g=\hat{\eta}, \forall g$. If the vector $\hat{\eta}$ is such that $\hat{\eta}_i>0, i \in \mathcal{X}$ and $\sum_i \hat{\eta}_i=1$, and
\begin{align}\label{condition}
0 < \sum_{g=1}^p \lambda_g q_{i_g,i}^g < 1, \qquad \forall (i,i_1,\ldots,i_p)\in \mathcal{X}^{p+1}\,,
\end{align}
then
\begin{align*}
\lim_{\ell \to \infty} \mathbb{P}(X_{t+\ell}=i|X_{t-1}=i_1,\ldots,X_{t-p}=i_p)=\hat{\eta}_{i}\,.
\end{align*}
\end{theorem}
The proof of the theorem is in Appendix \ref{app:B}. Notice that in this theorem we do not need to assume that the parameters $(\lambda_g,\boldsymbol{Q}^g)_{1\leq g \leq p}$ are between zero and one, but the probabilistic interpretation is guaranteed by the condition (\ref{condition}). Finally notice that the condition on $\hat{\eta}$ implies that $\forall g$ we can write $\boldsymbol{Q}^g=\boldsymbol{Q}+\tilde{\boldsymbol{Q}}^g$, where $\hat{\eta} \boldsymbol{Q} =\hat{\eta}$ and $\hat{\eta} \tilde{\boldsymbol{Q}}^g =0$. 

\subsection{Estimation}
Despite being parsimonious with respect to full Markov models, the MTDg parameters $\boldsymbol\theta=(\lambda_g,\boldsymbol{Q}^g)_{1\leq g \leq p}$ are difficult to estimate because they have to comply with the normalization constraints of transition matrices. In the literature many different estimation methods have been proposed~\cite{berchtold2002}, but in our paper we will focus on two specific methodologies: the maximum likelihood estimation (MLE) and the generalized method of moments (GMM). Let us introduce the details of these methods.

\subsubsection{Maximum likelihood estimation}
For a given data sequence with length $n$, $\lbrace X_t=x_t \rbrace_{t=1,\ldots,n}$, we define $(X_{t_1}^{t_2}=x_{t_1}^{t_2})$ as the sequence of events $(X_{t_1}=x_{t_1},X_{t_1+1}=x_{t_1+1},.....,X_{t_2}=x_{t_2})$ and $\mathbb{P}(X_1^p=x_1^p)$ is the joint distribution of $\lbrace X_t=x_t \rbrace_{t=1,\ldots,p}$. From the definition of MTDg models of order $p$, the likelihood function is
\begin{align}
L(\boldsymbol\theta)&=\mathbb{P}_{\boldsymbol\theta}(X_1^n=x_1^n)=\mathbb{P}(X_1^p=x_1^p)\mathbb{P}_{\boldsymbol\theta}(X_{p+1}^n=x_{p+1}^n|X_1^p=x_1^p) \nonumber \\
&=\mathbb{P}(X_1^p=x_1^p)\prod_{t=p+1}^n\left\lbrace\sum_{g=1}^p\lambda_g q^g_{x_{t-g},x_t}\right\rbrace\,, \label{eqn:l_function}
\end{align}
To estimate the parameters of MTDg model, we have excluded $\mathbb{P}(X_1^p=x_1^p)$ from the likelihood function. Therefore, the log-likelihood function that we consider is
\begin{equation}
\ell(\boldsymbol\theta)=\log{\mathbb{P}_{\boldsymbol\theta}(X_{p+1}^n=x_{p+1}^n|X_1^p=x_1^p)}=\sum_{t=p+1}^n\log{\left\lbrace\sum_{g=1}^p\lambda_g q^g_{x_{t-g},x_t}\right\rbrace}\,, \label{eqn:ll_function}
\end{equation}
where $\boldsymbol\theta=(\lambda_g,\boldsymbol{Q}^g)_{1\leq g \leq p}$ satisfies all the constraints of Eqs.~\ref{eqn:mtd_cond1}, \ref{eqn:mtd_cond2}, \ref{eqn:mtd_cond3} or \ref{condition}. Hence, the maximum likelihood estimation of the parameters $\hat{\boldsymbol\theta}=\left(\hat{\lambda}_g,\hat{\boldsymbol{Q}}^g\right)_{1\leq g \leq p}$ is the solution of the following constrained non-linear optimization problem:
\begin{align}
\left(\hat{\lambda}_g,\hat{\boldsymbol{Q}}^g\right)_{1\leq g \leq p}&=\underset{(\lambda_g,\boldsymbol{Q}^g)_{1\leq g \leq p}}{\operatorname{argmax}} \sum_{t=p+1}^n\log{\left\lbrace\sum_{g=1}^p\lambda_g q^g_{x_{t-g},x_t}\right\rbrace}\,, \nonumber \\
\mbox{s.t} \quad \sum_{g=1}^p \lambda_g&=1\,, \nonumber \\
\lambda_g &\geq 0, \qquad \forall g \in \lbrace 1,\ldots,p\rbrace \nonumber \\
q_{i,j}^g &\ge 0 \quad \mbox{and} \quad \sum_{j=1}^m q_{i,j}^g=1 \qquad \forall g \in \lbrace 1,\ldots,p\rbrace, \forall i,j\in \mathcal{X}\,.
\label{eqn:mtd_mle}
\end{align}

Clearly the solution of the previous optimization problem is very hard due to the high number of constraints. Berchtold~\cite{berchtold2001} proposes an efficient iterative process with the boundary adjustment in the MLE process which leads to a modification of the Newton's method. Under the constraints of Eq.~\ref{eqn:mtd_cond1} and \ref{eqn:mtd_cond2}, L\`{e}bre and Bourguignon in~\cite{lebre2008} introduce a hidden process for the coefficients of the MTDg and propose an Expectation-Maximization approach for the parameters estimation. Chen et al. in~\cite{chen2009} note that all the previous constraints can be rewritten in a box-constrained form, which is easier to handle.

\subsubsection{Generalized Method of Moments}
Raftery in~\cite{raftery1985} shows that the bivariate distributions of the MTD model satisfy a linear system of equations similar to the Yule-Walker equations. Here we extend this result to the MTDg case, i.e. when transition matrices $\boldsymbol{Q}^g$ differ at each lag $g$. In Appendix \ref{app:A} we prove the following proposition:
\begin{proposition}
  Suppose that a sequence of random variables $\left\lbrace X_t\right\rbrace_{t\in \mathbb{N}}$ taking values in the finite set $\mathcal{X}=\lbrace 1,\ldots,m\rbrace$ is defined by Eq.~\ref{eqn:mtdg} and is stationary. Let $\boldsymbol{B}(k)$ be a $m \times m$ matrix with elements
  \begin{align*}
    b_{i,j}^k=\mathbb{P}(X_t=i, X_{t+k}=j), \qquad i,j \in \mathcal{X}; k\in \mathbb{Z}
  \end{align*}
  and $\boldsymbol{B}(0)=\mbox{diag}(\hat{\eta}_1,\ldots,\hat{\eta}_m)$. Then
  \begin{align} \label{eqn:biv_system}
    \boldsymbol{B}(k)=\sum_{g=1}^p \lambda_g \boldsymbol{B}(k-g) \boldsymbol{Q}^g\,.
  \end{align}
\end{proposition}
The  system (\ref{eqn:biv_system}) consists in $m^2p$ different equations which can be employed as orthogonality conditions of the GMM applied to the MTDg model. These equations are not all independent, because the matrices of the bivariate distributions $\boldsymbol{B}(k)$ satisfy the usual normalization conditions. In fact, the rows and the columns of each matrix sum up to the corresponding unconditional probability, $\sum_j{b_{i,j}^k}=\hat{\eta}_i$ and $\sum_i{b_{i,j}^k}=\hat{\eta}_j$. By using these relations, the number of independent equations is reduced to $p(m^2-2m+1)$. The uniquenes of the solution of the system of Eq.~\ref{eqn:biv_system} requires that the number of independent parameters of the model has to be equal to the number of independent equations. 

\section{MTD for order flow and price impact}
\label{sec:mtdmicro}
We consider the joint dynamics of order flow and price changes in transaction time $t\in \mathbb{N}$. Each event is a transaction which has a positive sign ($\epsilon_t=+1$) if it is buyer initiated or negative ($\epsilon_t=-1$) if is seller initiated. For the price we distinguish two possibilities, namely that the trade changes the price ($\pi_t=C$) or not ($\pi_t=NC$). Notice that we are not considering the amplitude if the immediate price changes. For large tick stocks this is a minor problem, since price changes almost always of $\pm1$ tick, while for small tick stocks this is not true and we lose the information on the size of price change. In this paper we use the MTDg model to describe the sequence of signed events
\begin{equation*}
\left\lbrace (\epsilon_t,\pi_t) \right\rbrace_{t\in \mathbb{N}} \rightarrow \left\lbrace X_t \right\rbrace_{t\in \mathbb{N}},
\end{equation*}
hence the number of states of the model is $m=4$. The relation between the states of the model and the signed events is obtained with the arbitrary mapping
\begin{align*}
  \epsilon_t=-1,\pi_t=\mathrm{C} \quad &\rightarrow \quad X_t=1\,, \\
  \epsilon_t=-1,\pi_t=\mathrm{NC} \quad &\rightarrow \quad X_t=2\,, \\
  \epsilon_t=+1,\pi_t=\mathrm{NC} \quad &\rightarrow \quad X_t=3\,, \\
  \epsilon_t=+1,\pi_t=\mathrm{C} \quad &\rightarrow \quad X_t=4\,.
\end{align*}

The main quantity of interest is the cross and autocorrelation functions $C_{\pi_1,\pi_2}(\ell)$, already introduced in~\cite{eisler2012a,eisler2012b,taranto16}. Since
\begin{equation*}
  \hat{\eta}=\mathbb{P}(X_t)\equiv\mathbb{P}(\epsilon_t,\pi_t) \qquad \boldsymbol{B}(\ell)=\mathbb{P}(X_t;X_{t+\ell})\equiv\mathbb{P}(\epsilon_t,\pi_t;\epsilon_{t+\ell},\pi_{t+\ell})
\end{equation*}
these correlations
\begin{align*}
  C_{\pi_1,\pi_2}(\ell)&=\frac{\mathbb{E}[\epsilon_t I(\pi_t=\pi_1) \cdot \epsilon_{t+\ell} I(\pi_{t+\ell}=\pi_2)]}{\mathbb{P}(\pi_1)\mathbb{P}(\pi_2)} \\
  &=\sum_{\epsilon_{t}\epsilon_{t+\ell}}\sum_{\pi_t \pi_{t+\ell}}\frac{\epsilon_t I(\pi_t=\pi_1) \epsilon_{t+\ell} I(\pi_{t+\ell}=\pi_2) \mathbb{P}(\epsilon_t,\pi_t;\epsilon_{t+\ell},\pi_{t+\ell})}{P(\pi_1)P(\pi_2)} \,,
\end{align*}
where $I(\pi_t=\pi)$ is the indicator function, can be expressed in terms of $\hat{\eta}$ and $\boldsymbol{B}(\ell)$. For instance, for $\pi_t=\mathrm{NC}$ and $\pi_{t+\ell}=\mathrm{NC}$ the following relations hold
\begin{align*}
  \mathbb{P}(\mathrm{NC})&=\hat{\eta}_2+\hat{\eta}_3\,,\\
  C_{\mathrm{NC},\mathrm{NC}}(\ell)&=\frac{b_{2,2}(\ell)-b_{2,3}(\ell)-b_{3,2}(\ell)+b_{3,3}(\ell)}{(\hat{\eta}_2+\hat{\eta}_3)^2}\,.
\end{align*}

In the next two subsections we will estimate MTDg models on real financial data of the US markets. We will consider two different parametrizations and estimation methods. The first one, used as a benchmark case, is based on MLE and uses a parametrization which preserves the probabilistic interpretation of the mixture, i.e. it assumes that the parameters $(\lambda_g,\boldsymbol{Q}^g)_{1\leq g \leq p}$ are between zero and one. Moreover, in order to be able to apply MLE, we will impose a very strong structure of $(\lambda_g,\boldsymbol{Q}^g)_{1\leq g \leq p}$, reducing the number of parameters from $p(m^2-m)+p-1\sim 1,300$ for $p=100$ to $11$.  

In the second case we relax the constraint that $(\lambda_g,\boldsymbol{Q}^g)_{1\leq g \leq p}$ are between zero and one and we use GMM. We show that a suitable parametrization allows to reduce the estimation to the solution of a constrained linear system, which we prove to be a convex problem. This model is weakly constrained and we are able to estimate reliably $500$ parameters, improving significantly the performance of the model with respect to the benchmark case.  

\subsection{Strongly constrained MTDg model}
Estimation methods for the MTDg model proposed so far in literature have dealt with low order models. Unfortunately, our case requires the estimation of an high-order version of the model to capture the long-ranged dependence measured for the flow of trade signs. The log-likelihood function of Eq.~\ref{eqn:ll_function} is highly non-linear and the solution of the optimization problem could be very hard to find for large values of $p$.

{\bf Parametrization.} In order to reduce the number of parameters and to avoid non-linear constraints, we impose a functional form for the parameters which automatically satisfies all the constraints. For the $\lambda_g$ it is very natural to assume a power law scaling, $\lambda_g=N_\beta g^{-\beta}$, where $N_\beta^{-1}=\sum_{i=1}^p g^{-\beta}$. The reason behind this choice is that the values of $\lambda_g$ influence the correlations for large lags $\ell$, which empirically decay slowly with the lag. Another significant simplification of the problem can be achieved by assuming a buy/sell symmetry, which leads to the definition of centro-symmetric matrices $\boldsymbol{Q}^g$. This assumption leads to
\begin{equation*}
  q_{ij}^g=q_{m-i+1,m-j+1}^g, \quad \mbox{for } i,j=1,\ldots,m\,,
\end{equation*}
and for the first-order stationary distribution of the process
\begin{equation*}
  \hat{\eta}_i=\hat{\eta}_{m-i+1}, \quad \mbox{for } i=1,\ldots,m\,.
\end{equation*}
For instance, $q_{12}^g=q_{43}^g$ since the influence of a sell order price changing event at time $t-g$ on the probability of a sell order not price changing event at time $t$ is equal to the influence of a buy order price changing event at time $t-g$ on the probability of a buy order not price changing event at time $t$.

As mentioned above (see theorem \ref{teo1}), we consider matrices $\boldsymbol{Q}^g$  sharing the same left eigenvector with eigenvalue one. Writing $\boldsymbol{Q}^g=\boldsymbol{Q}+\tilde{\boldsymbol{Q}}^g$, we make the following strongly parametrized ansatz: 
\begin{align}
\boldsymbol{Q}=\begin{pmatrix}
B_1 & A_1 & A_1 & B_1 \\
B_2 & A_2 & A_2 & B_2 \\
B_2 & A_2 & A_2 & B_2 \\
B_1 & A_1 & A_1 & B_1
\end{pmatrix}, \qquad \boldsymbol{\tilde{Q}}^g=\begin{pmatrix}
-\mu_1 e^{-\alpha_{11}g} & -\nu_1 e^{-\alpha_{12} g} & \nu_1 e^{-\alpha_{12} g} & \mu_1 e^{-\alpha_{11}g} \\
\mu_2 e^{-\alpha_{21}g} & \nu_2 e^{-\alpha_{22} g} & -\nu_2 e^{-\alpha_{22} g} & -\mu_2 e^{-\alpha_{21}g} \\
-\mu_2 e^{-\alpha_{21}g} & -\nu_2 e^{-\alpha_{22} g} & \nu_2 e^{-\alpha_{22} g} & \mu_2 e^{-\alpha_{21}g} \\
\mu_1 e^{-\alpha_{11}g} & \nu_1 e^{-\alpha_{12} g} & -\nu_1 e^{-\alpha_{12} g} & -\mu_1 e^{-\alpha_{11}g}
\end{pmatrix}\,, 
\label{eqn:mtd_exp}
\end{align}
where $\alpha_{ij}\geq 0$. Imposing
\begin{align}
  A_1&=1/2-B_1, & A_2&=1/2-B_2\,, \nonumber \\
  0 \leq &B_1 \leq 1/2, & 0 \leq &B_2 \leq 1/2\,, \nonumber \\
  -B_1 \leq &\mu_1 \leq B_1, & -B_2 \leq &\mu_2 \leq B_2\,, \nonumber \\
  B_1-1/2 \leq &\nu_1 \leq 1/2-B_1, & B_2-1/2 \leq &\nu_2 \leq 1/2-B_2\,,
  \label{eqn:mtd_exp_cond}
\end{align}
we automatically satisfy all the constraints of the model. Moreover it is immediate to see that $\hat{\eta}\tilde{\boldsymbol{Q}}^g=0$, as required, where 
\begin{equation*}
  \hat{\eta}=\begin{pmatrix} \frac{B_2}{1-2B_1+2B_2}, \frac{1-2B_1}{2-4B_1+4B_2}, \frac{1-2B_1}{2-4B_1+4B_2}, \frac{B_2}{1-2B_1+2B_2}\end{pmatrix}\,,
\end{equation*}
Finally the parametrization in Eq.~\ref{eqn:mtd_exp} with the linear constraints of Eq.~\ref{eqn:mtd_exp_cond} guarantees that the matrices have the right normalization on the rows, $\sum_{j}q_{ij}^g=1, \forall g,i$ and $0<q_{ij}^g<1, \forall g,i,j$.

The intuition behind our choice is that the parameters $q_{ij}^g$ determine the correlations between the event $i$ at time $t$ and the event $j$ at time $t-g$. From the left panel in Figure 4~in~\cite{taranto16}, reporting the empirical correlations measured for the large tick stock Microsoft, we see a quite different behavior depending on the conditioning event. For example, the order flow correlations among non price-changing events is extremely persistent. This has motivated the choice of a power law decaying pre-factor $\lambda_g$. However, since $\lambda_g$ multiplies all entries of the matrices $\boldsymbol{Q}^g$ we need to include different decays in the $\tilde{\boldsymbol{Q}}^g$ matrices in order to reproduce the faster decay of the empirical correlations which involve price-changing events, and for this reason we have introduced the four exponential decay rates $\alpha_{ij}$ ($i,j=1,2$).  

The parameters of this model can be obtained via MLE. The optimization problem is non-trivial since the likelihood function is highly non-linear. However the dimensionality is low and, thanks to the parametrization, the constraints of the problem are linear inequalities. The total number of parameters is 11, $\boldsymbol{\theta}=\{\beta,B_i,\mu_i,\nu_i,\alpha_{ij}\}$ with $i,j=1,2$. 

\begin{figure}[p]
\centering
\includegraphics[width=0.8\columnwidth]{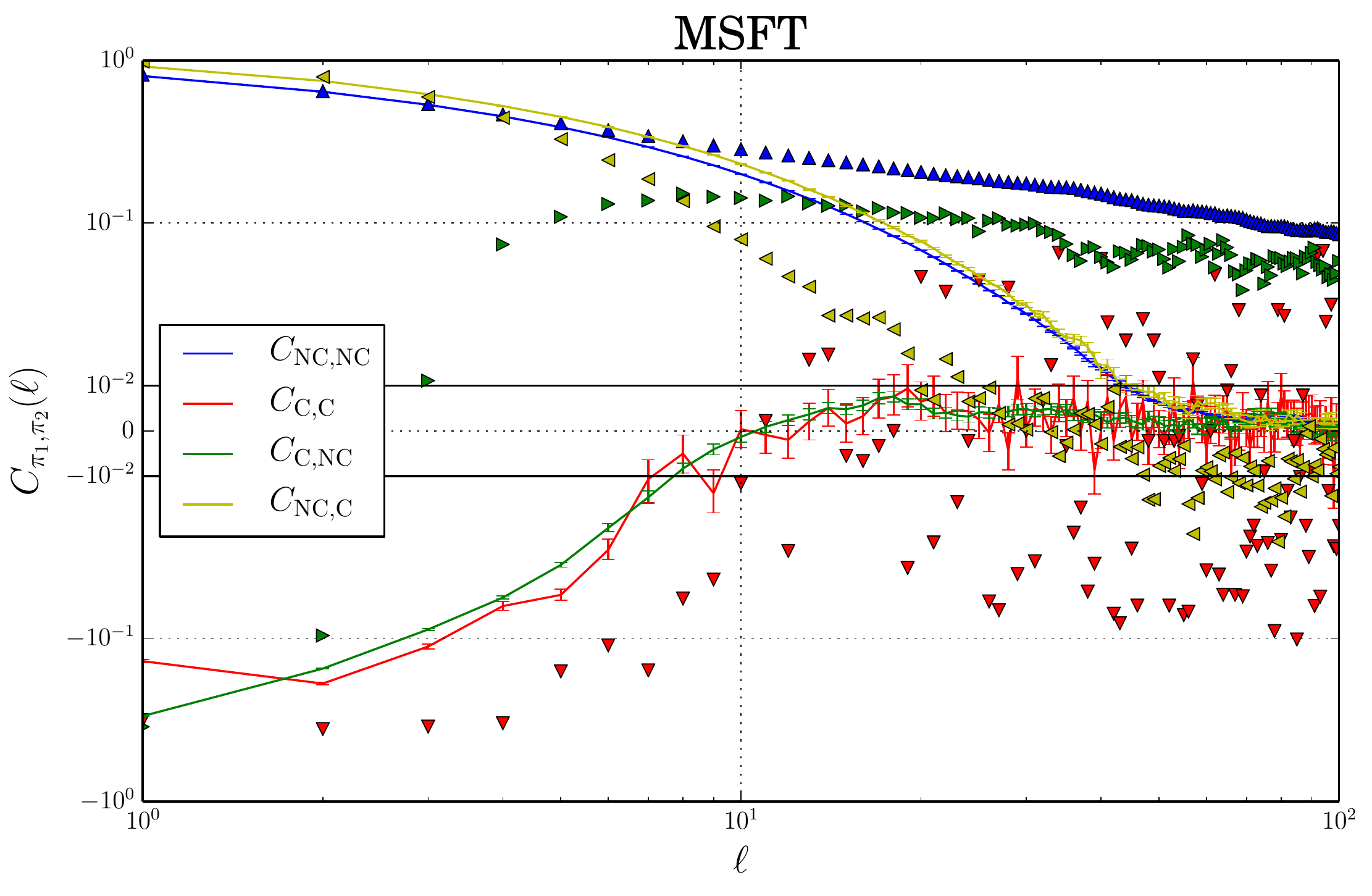}
\caption{\textit{MLE calibration of the strongly constrained MTDg}. Comparison between the auto and cross-correlation functions $C_{\pi_1,\pi_2}(\ell)$  of signed events from a simulation of the MTDg(100) model estimated on MSFT data (solid lines) and the empirical curves (triangles). The error bars correspond to one standard deviation. Estimated parameter values are $\beta=2.38$, $B_1=0.018$, $B_2=0.40$, $\mu_1=0.018$,  $\alpha_{11}=0.0$, $\nu_1=0.48$, $\alpha_{12}= 0.47$, $\mu_2=0.04$, $\alpha_{21}=0.003$, $\nu_2=0.42$, and $\alpha_{22}=0.0$. The scale for values close to zero and bounded by horizontal solid lines is linear, whereas outside this region the scale is logarithmic.}
\label{fig:sim_mtd_exp_eps_msft}
\includegraphics[width=0.8\columnwidth]{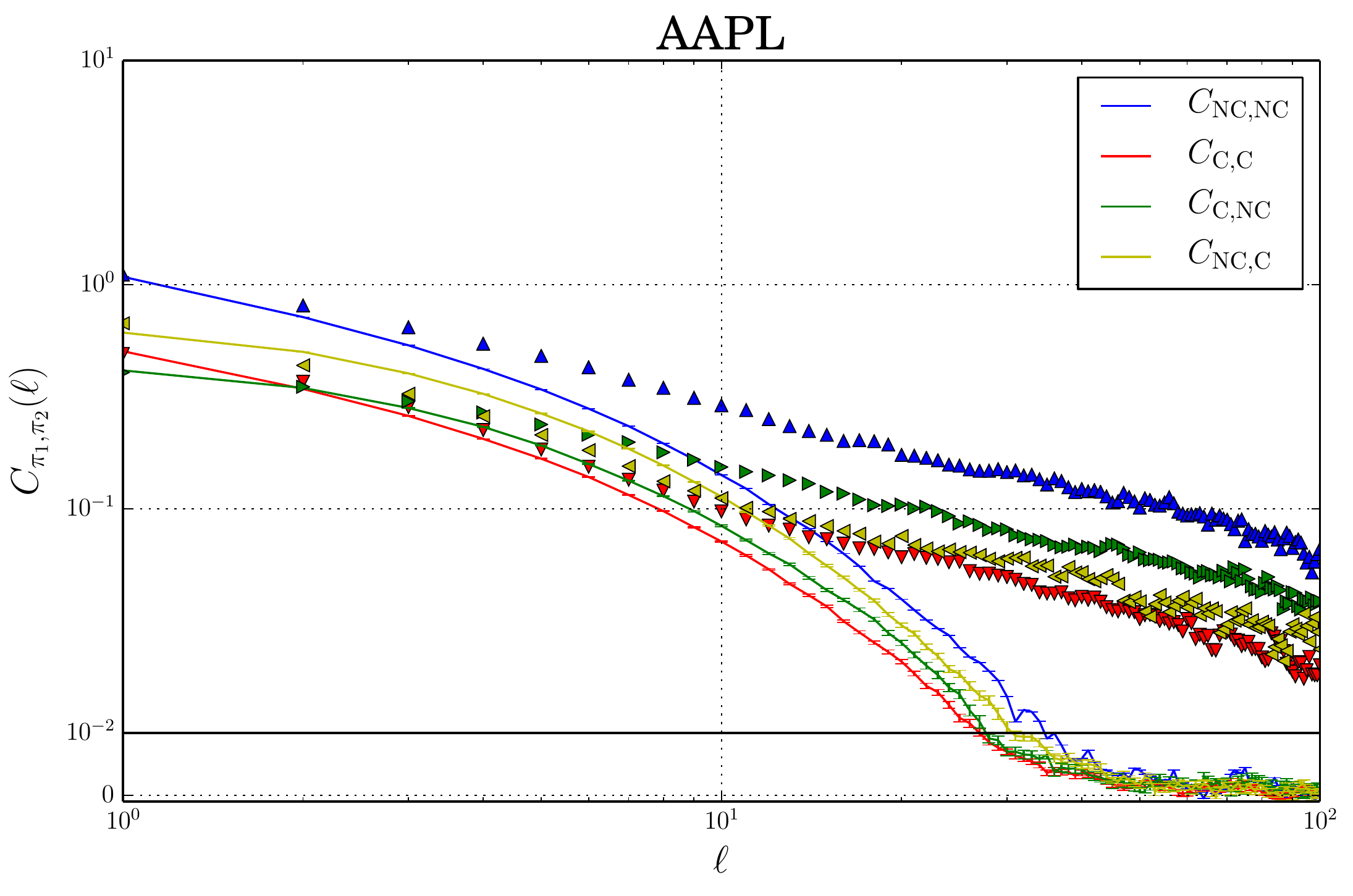}
\caption{\textit{MLE calibration of the strongly constrained MTDg}. Comparison between the auto and cross-correlation functions $C_{\pi_1,\pi_2}(\ell)$  of signed events from a simulation of the MTDg(100) model estimated on AAPL data (solid lines) and the empirical curves (triangles). The error bars correspond to one standard deviation. Estimated parameter values are $\beta=2.21$, $B_1=0.38$, $B_2=0.01$, $\mu_1=-0.22$, $\alpha_{11}=0.0$, $\nu_1=-0.07$, $\alpha_{12}=0.0$, $\mu_2=0.27$, $\alpha_{21}=0.043$, $\nu_2=0.21$, and $\alpha_{22}=0.0$. The scale for values close to zero and bounded by horizontal solid lines is linear, whereas outside this region the scale is logarithmic.}
\label{fig:sim_mtd_exp_eps_aapl}
\end{figure}

{\bf Results.}  In Fig.~\ref{fig:sim_mtd_exp_eps_msft} and \ref{fig:sim_mtd_exp_eps_aapl}  we plot the correlation functions computed from a Monte Carlo simulation of the MTDg(100) model with parameter values obtained from MLE on Microsoft (MSFT) and Apple (AAPL) data (details about the data set are given in Section 4.1 of the companion paper~\cite{taranto16}). More precisely, we compare the auto and cross-correlations $C_{\pi_1,\pi_2}(\ell)$ for price-changing and non price-changing events with the empirical ones. As can be noted, for small tick stocks the model can reproduce the structure of the correlations for short time scales, but not their persistence. For large tick stocks the persistence of the empirical correlations is not well reproduced either, and the structure of the correlations predicted by MTD for small lags is not rich enough to fit the empirical data. The lack in the persistence of the simulated correlations can be explained by the fact that the optimized exponent $\beta$ is too high. Also for the case of large tick stocks we conclude that the functional forms assumed for the matrices $\boldsymbol{Q}^g$ is not flexible enough to reproduce the different speed of decays of the empirical correlations. Nonetheless, the advantage of this modeling assumption is that the estimation process is extremely fast even for higher order models. In the next subsection we will explore a better alternative.

\subsection{Weakly constrained MTDg model}
\label{sec:weakly}
{\bf Model definition.} Here we introduce the main methodological innovation of this paper, namely a parametrization of the MTDg model which can be estimated with GMM even when the number of parameters is very large. To motivate it, let us consider the DAR(p) process with $m$ states (employed for example in~\cite{taranto2014} as a model for the order flow) \footnote{The case $m=2$ considered in Ref.~\cite{taranto16} corresponds to a MTD(p) model with 
transition matrices that are the same for all $g$, $\boldsymbol{Q}^g \equiv \boldsymbol{Q}$ and
\begin{equation*}
  \boldsymbol{Q}=\begin{pmatrix}
  \rho & 1-\rho \\
  1-\rho & \rho
  \end{pmatrix}\,. 
\end{equation*}
In the stationary condition the two states are equiprobable, as can be verified solving the left eigenvalue problem $\hat{\eta} \boldsymbol{Q}=\hat{\eta}$.}. 
The model can be seen as a particular case of the MTD(p) model, where the transition matrices are the same for all $g$, $\boldsymbol{Q}^g \equiv \boldsymbol{Q}$ and such that 
\begin{equation*}
  \boldsymbol{Q}=1^T\hat{\eta}+\varphi(\mathbb{I}-1^T\hat{\eta})\,,
\end{equation*}
where $1$ is a row of $m$ ones and the parameter $\varphi$ ranges between zero and one. The left eigenvector of  $\boldsymbol{Q}$ corresponding to the eigenvalue $1$ is $\hat{\eta}$, since it belongs to the kernel of $\mathbb{I}-1^T\hat{\eta}$.


Following the same idea, we introduce MTD(p) models where  
\begin{align}
\boldsymbol{Q}^g&=1^T\hat{\eta}+\boldsymbol{\tilde{Q}}^g\
\end{align}
and $\hat{\eta} \boldsymbol{\tilde{Q}}^g=0$. Moreover normalization of $\boldsymbol{Q}^g$ imposes that each row of $\boldsymbol{\tilde{Q}}^g$ sums to zero, hence these matrices will have negative elements. As in Theorem \ref{teo1}, all the $\boldsymbol{Q}^g$ share the same left eigenvector $\hat{\eta}$ with eigenvalue $1$. It is easy to show that the conditional probabilities of this model can be written as
\begin{align}
  \mathbb{P}(X_t=i|X_{t-1}=i_1,\ldots,X_{t-p}=i_p) = \hat{\eta}_{i}+\sum_{g=1}^p a_{i_g,i}^g\,,
  \label{eqn:smart_mtdg}
\end{align}
where $a_{i_g, i}^g \equiv \lambda_g (\boldsymbol{\tilde{Q}}^g)_{i_g,i}$. Thus the matrices $\boldsymbol{A}^g \equiv \lambda_g\boldsymbol{\tilde{Q}}^g$ describe the deviations of the $p-$order transition probability from the stationary value given by $\hat{\eta}$. Finally, as shown in Appendix \ref{app:C}, the  system of equations of Theorem \ref{teo1} for this model is 
\begin{align}
  \boldsymbol{B}(k)-\hat{\eta}^T\hat{\eta}=\sum_{g=1}^p \boldsymbol{B}(k-g) \boldsymbol{A}^g.
  \label{eqn:smart_biv_system}
\end{align}
 
This linear system can be used to estimate the model, i.e. the matrices $\boldsymbol{A}^g$, from the knowledge of the stationary probabilities $\hat{\eta}$ and the bivariate distributions $\boldsymbol{B}(k)$. There are however two technical problems: 
\begin{itemize}
\item The estimated model might not have a probabilistic interpretation, i.e. the estimated model might generate transition probabilities larger than one or smaller than zero; 
\item The solution of Eq.~\ref{eqn:smart_biv_system} gives the matrix $\boldsymbol{A}^g$, while one might need  $\lambda_g$ and $(\boldsymbol{\tilde{Q}}^g)$ separately, thus the identifiability problem must be solved by fixing arbitrarily one parameter. Note however that the dynamics of the model is independent from this choice. 
\end{itemize}
In the following we will tackle these points.

In order to have a well defined probabilistic model, and to be able to use Theorem \ref{teo1} which guarantees the existence and uniqueness of the solution, it must also hold that    
\begin{align*}
  0 < \hat{\eta}_{i}+\sum_{g=1}^p a_{i_g,i}^g < 1, \qquad \forall (i,i_1,\ldots,i_p)\in \mathcal{X}^{p+1}\,,
\end{align*}
which corresponds to $2m^{p+1}$ constraints. 
Clearly, in practical applications it is impossible to handle the previous number of conditions, but we can satisfy all of them imposing the necessary and sufficient conditions
\begin{align}
  \hat{\eta}_{i}+\sum_{g=1}^p \max_{i_g} \left(a_{i_g,i}^g \right) &< 1, \qquad \forall i \in \mathcal{X} 
  \label{eqn:smart_mtdg_comb1} \\
  \hat{\eta}_{i}+\sum_{g=1}^p \min_{i_g} \left(a_{i_g,i}^g \right) &> 0, \qquad \forall i \in \mathcal{X}
  \label{eqn:smart_mtdg_comb2}
\end{align}
which are only $2m$ inequality constraints. Under these conditions the process is well defined and possesses a unique stationary solution (Theorem \ref{teo1}) and the estimation of the model can be performed solving the optimization program 

\begin{align}
  \hat{q}&=\underset{{\bf q} \in \mathbb{R}^{p(m^2-2m+1)}}{\operatorname{argmin}} \left\Vert {\bf d}-\boldsymbol{K}\cdot {\bf q} \right\Vert^2 \nonumber \\
  \mbox{s.t.} \qquad &\hat{\eta}_{i}+\sum_{g=1}^p \max_{i_g} \left(a_{i_g,i}^g \right) < 1, \qquad \forall i \in \mathcal{X}  \nonumber \\
  &\hat{\eta}_{i}+\sum_{g=1}^p \min_{i_g} \left(a_{i_g,i}^g \right) > 0, \qquad \forall i \in \mathcal{X} \label{eqn:mtd_yw_min_contrained}
\end{align}
where the elements of the $p(m^2-2m+1)$-dimensional vector $\bf d$ correspond to left hand side of Eq.~\ref{eqn:smart_biv_system}, namely
$${\bf d}=(\overline b_{1,1}^1,\ldots,\overline b_{1,m-1}^1,\ldots,\overline b_{m-1,1}^1,\ldots,\overline b_{m-1,m-1}^1,\ldots,\overline b_{1,1}^p,\ldots,\overline b_{1,m-1}^p,\ldots,\overline b_{m-1,1}^p,\ldots,\overline b_{m-1,m-1}^p)
$$
with 
$$
\overline b_{i,j}^k=b_{i,j}^k-\hat{\eta}_i\hat{\eta}_j ,
$$
 the vector $\bf q$ corresponds to the parameters of the model $\lambda_g \tilde{q}_{i,j}^g$
 $$
 {\bf q}=(a_{1,1}^1,\ldots,a_{1,m-1}^1,\ldots,a_{m-1,1}^1,\ldots,a_{m-1,m-1}^1,\ldots,a_{1,1}^p,\ldots,a_{1,m-1}^p,\ldots,a_{m-1,1}^p,\ldots,a_{m-1,m-1}^p)
 $$
and the elements of the matrix $\boldsymbol{K}$ are linear combinations of $b_{i,j}^k$, according to Eq.~\ref{eqn:smart_biv_system} (we do not report the matrix since its form is not transparent). 

The reason for the choice of the constraints in Eq.~\ref{eqn:mtd_yw_min_contrained} is that we prove in Appendix \ref{app:D} the following proposition:
\begin{proposition}
If $\boldsymbol{K}$ is not singular, the optimization program of Eq.~(\ref{eqn:mtd_yw_min_contrained}) is strictly convex in $\mathbb{R}^{p(m^2-2m+1)}$.  
\end{proposition}
Therefore if a local minimum exists, then it is a global minimum. The convexity property solves the issue of the high dimensionality of the problem and the model can be estimated also for large order $p$.

{\bf Application to order flow and impact.} We now consider the application of the above described MTDg model to the $m=4$ process describing jointly the order flow and the price changes.  As done in the previous section, we reduce the dimensionality of the system by exploiting the buy/sell symmetry, which leads to centrosymmetric $\hat{\eta}$ and $\boldsymbol{B}(k)$. In fact, for $m=4$ we have that
\begin{equation*}
  b_{i,j}^k=b_{m-i+1,m-j+1}^k, 
\end{equation*}
and for the stationary distribution 
\begin{equation*}
  \hat{\eta}_i=\hat{\eta}_{m-i+1}, \quad \mbox{for } i=1,\ldots,m\,.
\end{equation*}
The buy/sell symmetry and the normalization of matrices $\boldsymbol{B}(k)$ reduces the number of independent variables in $\boldsymbol{B}(k)$ to $5p$, $5$ for each lag $k$. Thus, we have that
\scriptsize
\begin{align*}
  \boldsymbol{B}(k)=\begin{pmatrix}
    b_{1,1}^k & b_{1,2}^k & \hat{\eta}_2-b_{1,2}^k-b_{2,2}^k-b_{3,2}^k & \hat{\eta}_1-\hat{\eta}_2+b_{2,2}^k+b_{3,2}^k-b_{1,1}^k \\
    b_{2,1}^k & b_{2,2}^k & b_{3,2}^k & \hat{\eta}_2-b_{2,1}^k-b_{2,2}^k-b_{3,2}^k \\
    \hat{\eta}_2-b_{2,1}^k-b_{2,2}^k-b_{3,2}^k & b_{3,2}^k & b_{2,2}^k & b_{2,1}^k \\
    \hat{\eta}_1-\hat{\eta}_2+b_{2,2}^k+b_{3,2}^k-b_{1,1}^k & \hat{\eta}_2-b_{1,2}^k-b_{2,2}^k-b_{3,2}^k & b_{1,2}^k & b_{1,1}^k
  \end{pmatrix}\,. 
\end{align*}
\normalsize

In order to find a solution of the problem of Eq.~\ref{eqn:mtd_yw_min_contrained}, we assume that the imposed centrosymmetry of $\boldsymbol{B}(k)$ and $\hat{\eta}$ does not change the rank of the matrix $\boldsymbol{K}$. In this case the solution is unique and it is easy to show that also $\boldsymbol{\tilde Q}^g$ must be centrosymmetric, as 
\scriptsize
\begin{align}
  \boldsymbol{\tilde{Q}}^g&=\begin{pmatrix}
    \tilde{q}_{1,1}^g & \tilde{q}_{1,2}^g & -\tilde{q}_{1,2}^g-c_2(\tilde{q}_{2,2}^g+\tilde{q}_{2,3}^g) & -\tilde{q}_{1,1}^g+c_2(\tilde{q}_{2,2}^g+\tilde{q}_{2,3}^g) \\
    \tilde{q}_{2,1}^g & \tilde{q}_{2,2}^g & \tilde{q}_{2,3}^g & -\tilde{q}_{2,1}^g-\tilde{q}_{2,2}^g-\tilde{q}_{2,3}^g \\
    -\tilde{q}_{2,1}^g-\tilde{q}_{2,2}^g-\tilde{q}_{2,3}^g & \tilde{q}_{2,3}^g & \tilde{q}_{2,2}^g & \tilde{q}_{2,1}^g \\
    -\tilde{q}_{1,1}^g+c_2(\tilde{q}_{2,2}^g+\tilde{q}_{2,3}^g) & -\tilde{q}_{1,2}^g-c_2(\tilde{q}_{2,2}^g+\tilde{q}_{2,3}^g) & \tilde{q}_{1,2}^g & \tilde{q}_{1,1}^g
  \end{pmatrix}\,,
  \label{eqn:smart_mtdg_model}
\end{align}
\normalsize
where $c_2=\hat{\eta}_2/\hat{\eta}_1$. With this definition the number of independent parameters in $\boldsymbol{Q}^g$ is also equal to 5 for each $g$. 

We can now solve the system of Eq.~\ref{eqn:mtd_yw_min_contrained} whose unknowns are the components of the matrix $\boldsymbol{A}^g$. This way we obtain the value of the products $\lambda_g \tilde{q}_{i,j}^g$, but not the value of the components $\lambda_g$ and $\tilde{q}_{i,j}^g$ separately. For this reason we impose that one of the five components among $\tilde{q}_{1,1}^g$, $\tilde{q}_{1,2}^g$, $\tilde{q}_{2,1}^g$, $\tilde{q}_{2,2}^g$, and $\tilde{q}_{2,3}^g$ is independent of the lag $g$. We arbitrarily fix $\tilde{q}_{2,1}^g\equiv \tilde{q}_{2,1}$. We are left with 4$p$ free parameters from $\boldsymbol{\tilde Q}^g$ (4 for each $g$), $p-1$ parameters from $\lambda_g$ and $\tilde{q}_{21}$. In total we have $5p$ free parameters, which is exactly the same number of independent components $b_{i,j}^k$. 
The values of the products $\lambda_g \tilde{q}_{i,j}^g$ define the MTDg model.  Different choices of $\tilde{q}_{i,j}^g=\tilde{q}_{i,j}$ give  different factorizations, but lead to the same high-order Markov chain. The arbitrariness of the choice is an evidence of the well known identifiability problem of all mixture models.

In the literature there exist many algorithms which solve iteratively the constrained optimization problem of Eq.~\ref{eqn:mtd_yw_min_contrained}. A widely used class belongs to the Sequential Quadratic Programming (SQP) family~\cite{boggs1995}. However, an issue of our optimization is that constraints are non-smooth functions, which is a necessary condition required by the usual SQP algorithms. In a recent paper, Curtis and co-authors~\cite{curtis2012} have proposed the Sequential Quadratic Programming Gradient Sampling algorithm (SQP-GS), which can be applied to non-smooth, non-linear objective and constraint functions. We have implemented this algorithm in order to solve our optimization problem.

{\bf Results.} We estimated the above MTDg(100) model on MSFT and AAPL. Before showing the results, we mention that we have estimated also the model by using Eq.~(\ref{eqn:smart_biv_system}) {\it without} the additional constraints  of Eqs.~\ref{eqn:smart_mtdg_comb1} and \ref{eqn:smart_mtdg_comb2}. We found negative transition probabilities, indicating the importance to impose the constraints to have meaningful model estimation (and guarantee of existence and uniqueness of the solution).

\begin{figure}[p]
  \centering
  \includegraphics[width=0.8\columnwidth]{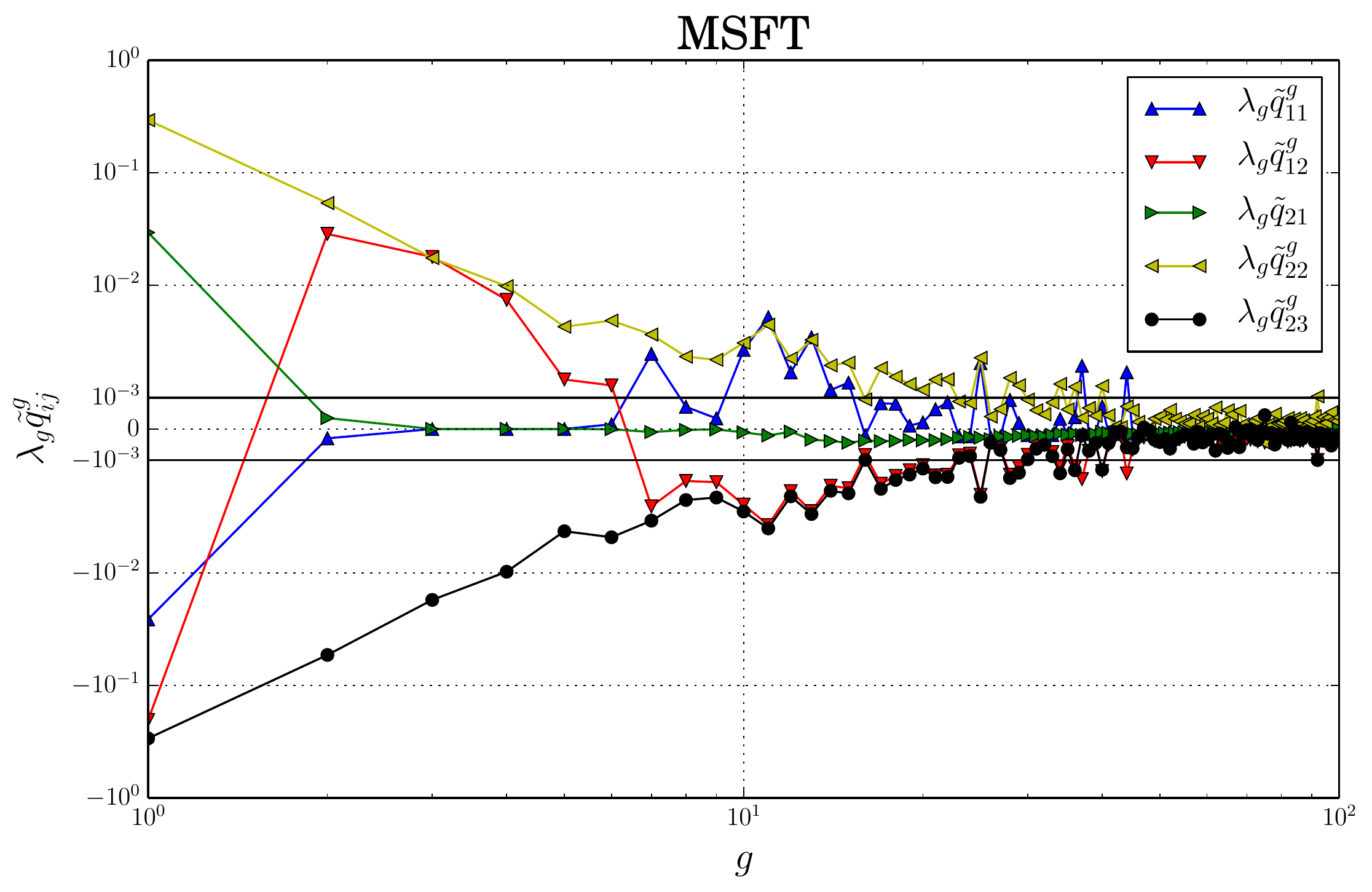}
  \caption{\textit{GMM calibration of the weakly constrained MTDg}. Plot of the parameters $a_{i,j}^g$ solution of the optimization problem of Eq.~\ref{eqn:mtd_yw_min_contrained} for an MTDg of order $p=100$ model estimated from MSFT data. The scale for values close to zero and bounded by horizontal solid lines is linear, whereas outside this region the scale is logarithmic.}
  \label{fig:mtd_yw_min_abscon_lambq_msft}
  \includegraphics[width=0.8\columnwidth]{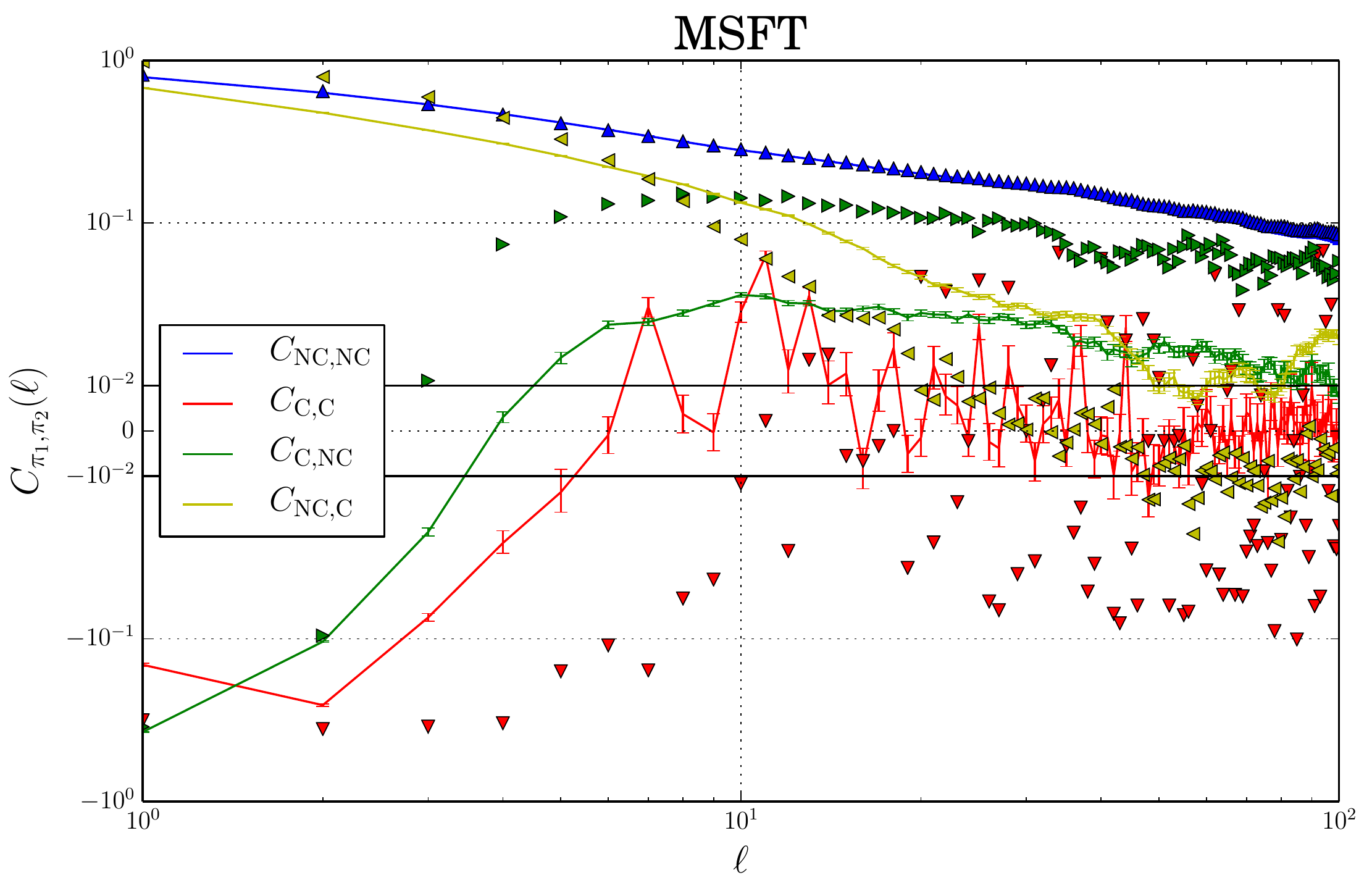}
  \caption{\textit{GMM calibration of the weakly constrained MTDg}. Comparison between the auto and cross-correlation functions $C_{\pi_1,\pi_2}(\ell)$ of signed events from a simulation of the MTDg(100) model estimated on MSFT data (triangles) and the empirical curves (solid lines). The error bars correspond to one standard deviation. The scale for values close to zero and bounded by horizontal solid lines is linear, whereas outside this region the scale is logarithmic.
  }
  \label{fig:sim_mtd_yw_min_abscon_msft}
\end{figure}

\begin{figure}[p]
  \centering
  \includegraphics[width=0.8\columnwidth]{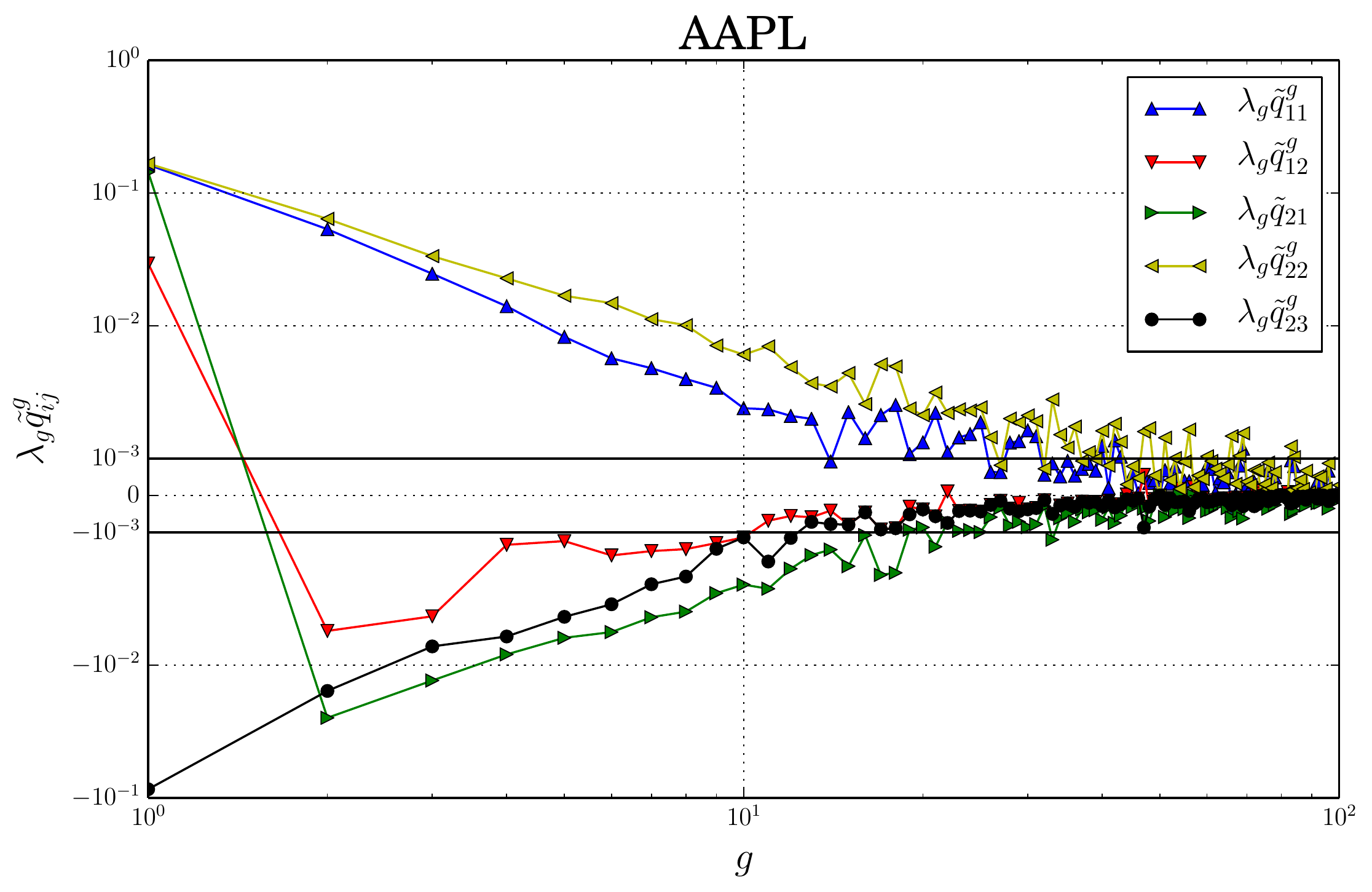}
  \caption{\textit{GMM calibration of the weakly constrained MTDg}. Plot of the parameters $a_{i,j}^g$ solution of the optimization problem of Eq.~\ref{eqn:mtd_yw_min_contrained} for an MTDg of order $p=100$ model estimated from AAPL data. The scale for values close to zero and bounded by horizontal solid lines is linear, whereas outside this region the scale is logarithmic.}
  \label{fig:mtd_yw_min_abscon_lambq_aapl}
  \includegraphics[width=0.8\columnwidth]{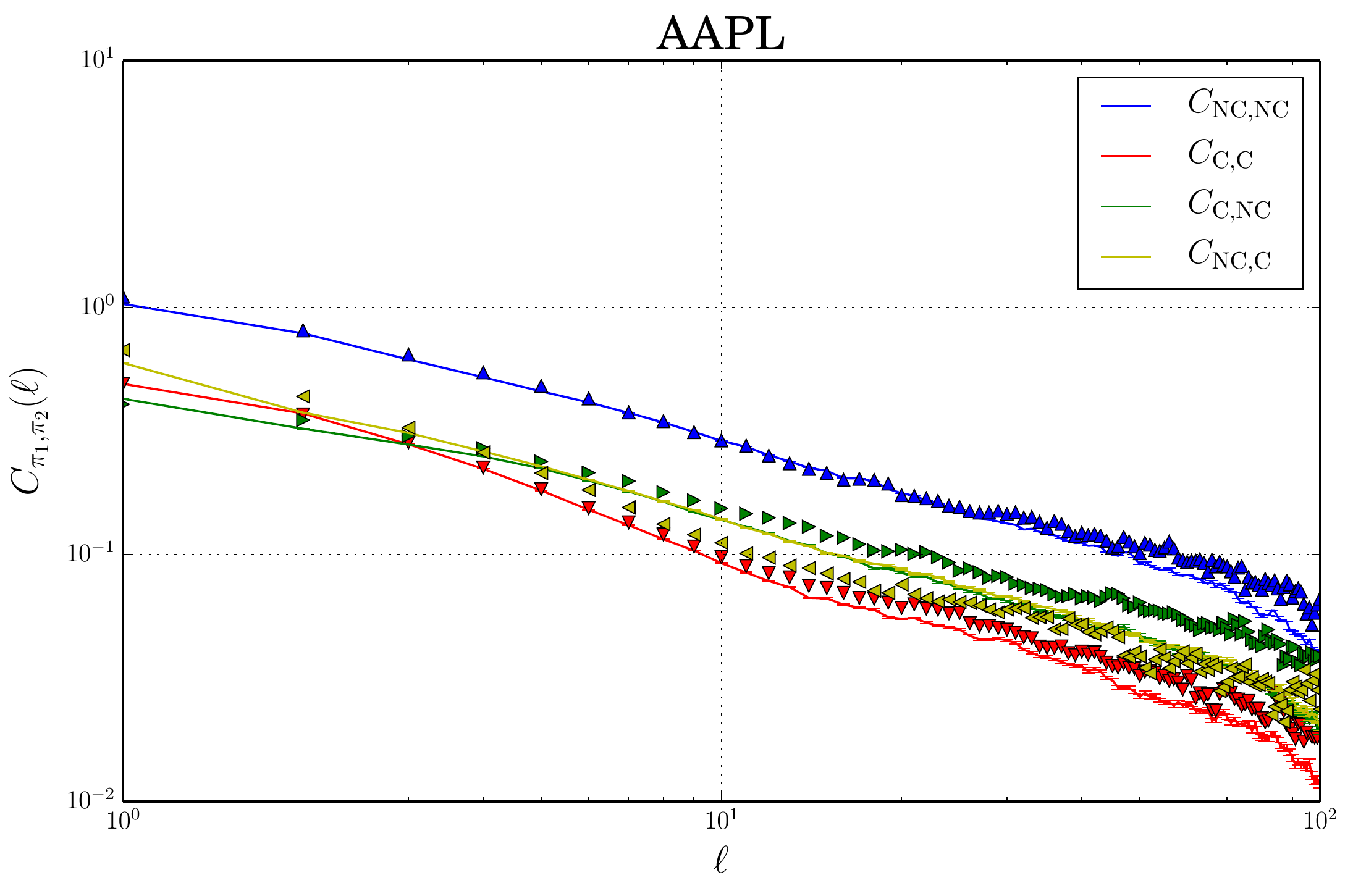}
  \caption{\textit{GMM calibration of the weakly constrained MTDg}. Comparison between the auto and cross-correlation functions $C_{\pi_1,\pi_2}(\ell)$ of signed events from a simulation of the MTDg(100) model estimated on AAPL data (triangles) and the empirical curves (solid lines). The error bars correspond to one standard deviation.}
  \label{fig:sim_mtd_yw_min_abscon_aapl}
\end{figure}

We now turn to the correctly constrained model. Fig.~\ref{fig:mtd_yw_min_abscon_lambq_msft} and \ref{fig:mtd_yw_min_abscon_lambq_aapl} show  the estimation of $\lambda_q\tilde q_{i,j}^g$ for MSFT and AAPL. Despite the large number of estimated parameters, they turn out to be only moderately noisy. Moreover it is interesting to note that negative values of $\lambda_q\tilde q_{i,j}^g$ are present, even if, by construction, the transition probabilities of the model are well defined in $[0,1]$. Clearly the estimation shows that the probabilistic mixture discussed at the beginning is, perhaps meaningfully, not suitable for the present data.  

Fig.~\ref{fig:sim_mtd_yw_min_abscon_msft} and \ref{fig:sim_mtd_yw_min_abscon_aapl} show correlation functions $C_{\pi_1,\pi_2}(\ell)$ of signed events computed from a Monte Carlo simulation of the calibrated model and compared with real data. As can be noted, for small tick stocks we have significantly improved the results of Fig.~\ref{fig:sim_mtd_exp_eps_aapl}. Compared with the benchmark, the new estimation method reproduces the high persistence of the correlations of order signs independently from the conditioning events. In the case of the large tick stocks, whose correlations present an highly non-trivial structure, the GMM methodology greatly improves the results with respect to Fig.~\ref{fig:sim_mtd_exp_eps_msft}. In particular, the high persistence of non price-changing events is very well reproduced. Moreover, the $C_{\mathrm{NC},\mathrm{C}}(\ell)$ curve decays faster as compared to the previous estimation method, and is thus closer to data.

\subsection{Large tick stock signature plot}
Another way to assess the quality of the MTDg model is to analyse how well it describes the volatility of prices. As noted above and in~\cite{taranto16},  the 
impact of a price changing event is nearly price independent for large tick stocks (within the TIM2 model). This means that the signature plot is simply given by:
\begin{equation}
  D^{\text{TIM2}}(\ell) \approx D_\mathrm{LF} + G_\mathrm{C}(1)^2 \mathbb{P}(\mathrm{C})+2 \frac{G_\mathrm{C}(1)^2}{\ell} \sum_{0 \leq n < m <\ell}\mathbb{P}(\mathrm{C})^2 C_{\mathrm{C}, \mathrm{C}}(m-n)\,,
  \label{eqn:sign_plot_mtd}
\end{equation}
which is completely determined by the correlation function $C_{\mathrm{C}, \mathrm{C}}(\ell)$ (once the value of $G_\mathrm{C}(1)$ has been estimated). This correlation function is, as presented above, only approximately reproduced by the MTDg model, although it is calibrated to minimize the distance to all 
$C_{\pi, \pi'}(\ell)$. In the context of financial applications, it is therefore interesting to replot the difference between the MTDg $C_{\mathrm{C}, \mathrm{C}}(\ell)$ and empirical data in terms of the signature plot $D^{\text{TIM2}}(\ell)$, which involves the integral of the correlation function. 

\begin{figure}[t]
  \centering
  \includegraphics[width=0.8\columnwidth]{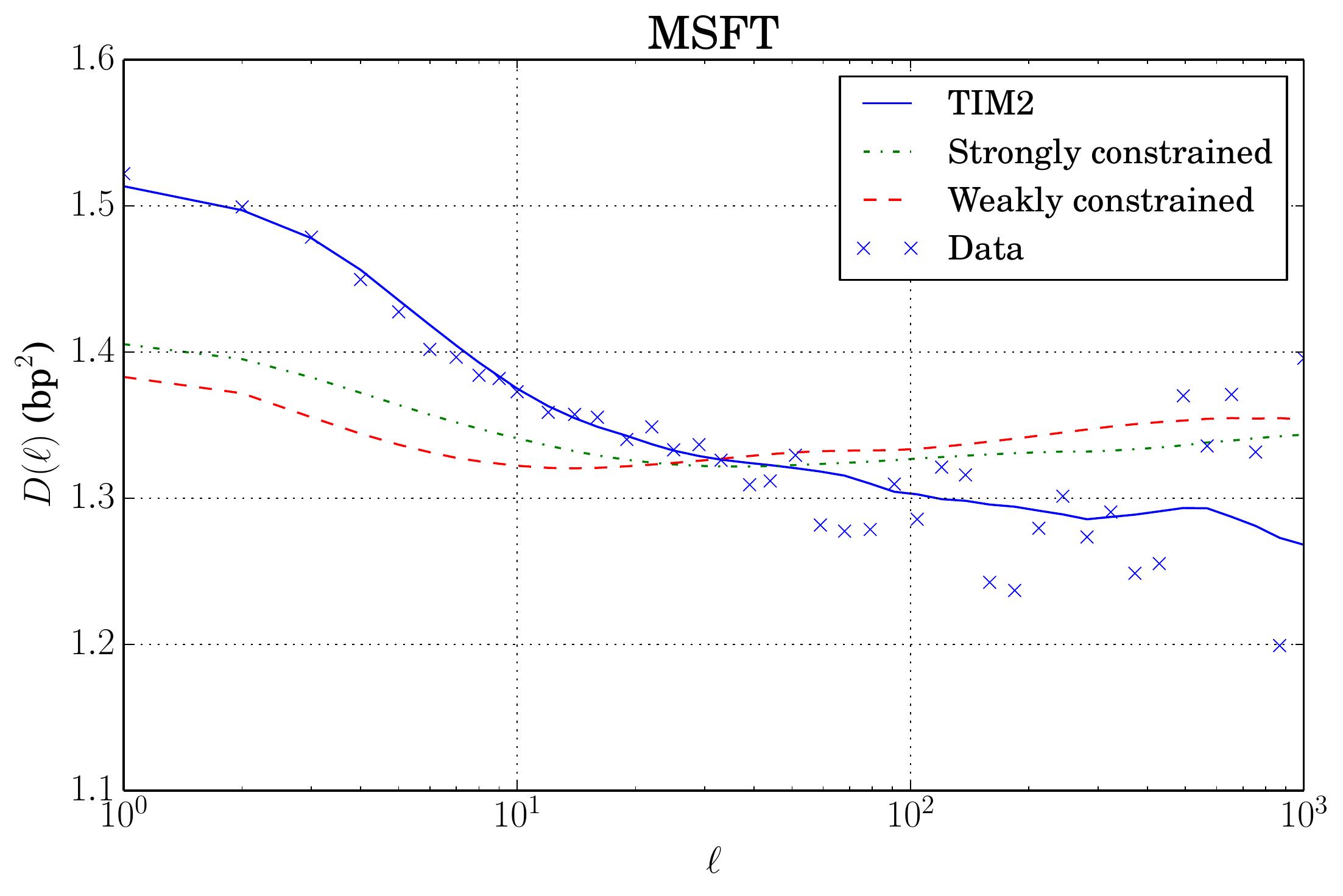}
  \caption{Signature plot for MSFT data: Empirical data (crosses), weakly constrained (GMM) MTDg(100) model with $D_\mathrm{LF}=0.41$ (dashed line), strongly constrained (MLE) MTDg(100) model with $D_\mathrm{LF}=0.43$ (dashed-dotted line), and the theoretical prediction of the calibrated TIM2 model~\cite{taranto16}.}
  \label{fig:signature_plot_msft}
\end{figure}

In Fig.~\ref{fig:signature_plot_msft} we show the curves corresponding to Eq.~\ref{eqn:sign_plot_mtd} for the strongly and weakly constrained versions of the MTDg model proposed above, where the extra fitting parameter $D_\mathrm{LF}$ is optimized with OLS in order to minimize the distance between the empirical and the theoretical curves of the model. We see that in terms of the signature plot of the model, the weakly constrained and strongly constrained MTDg fare nearly equally well. We also show the predictions of the TIM2 model that uses the empirical $C_{\mathrm{C}, \mathrm{C}}(\ell)$; the nearly perfect fit
in this case is a consequence of the fact that $G_\mathrm{C}(\ell) \approx G_\mathrm{C}(1)$ for large tick stocks. 

Note that the TIM2 price process is strictly diffusive only if the quantity $D^{\text{TIM2}}(\ell+1)(\ell+1)-D^{\text{TIM2}}(\ell)\ell$ is a constant independent from $\ell$. In fact, we have that
\begin{align*}
  D^{\text{TIM2}}(\ell+1)(\ell+1)-D^{\text{TIM2}}(\ell)\ell=D_\mathrm{LF} + G_\mathrm{C}(1)^2 \mathbb{P}(\mathrm{C})+2 G_\mathrm{C}(1)^2 \mathbb{P}(\mathrm{C})^2 \sum_{0 < n \leq \ell} C_{\mathrm{C}, \mathrm{C}}(n)\,,
\end{align*}
which means that the price process becomes diffusive for $\ell> \ell^*$ only if $C_{\mathrm{C}, \mathrm{C}}(\ell> \ell^*)=0$. 
Figs~\ref{fig:sim_mtd_exp_eps_msft} and \ref{fig:signature_plot_msft} suggest that this is indeed the case for $\ell^* \approx 10$.

\section{Out-of-sample analysis}
\label{sec:out}
In the previous sections we have presented two MTDg models -- strongly and weakly constrained -- and discussed two alternative estimation methodologies based on MLE and GMM. Since they differ both in the number of parameters and in estimation efficiency, it is important to compare their performances testing the predictive power of the models in an out-of-sample analysis. We consider as a measure of the performance the expected prediction error (EPE) defined as
\begin{equation*}
  \mathrm{EPE}(\boldsymbol\theta)=\mathbb{E}[L(X_t,\hat{X}_t^{\boldsymbol\theta})]\,,
\end{equation*}
where $X_t$ is the observed process, $\hat{X}_t^{\boldsymbol\theta}$ is the predictor of $X_t$ based on the model with parameter set $\boldsymbol\theta$, and the $p$ past observations of the process $X_t$. 
As common in the literature for categorical data,
we use as loss function the log-likelihood $L(X_t,\hat{X}_t^{\boldsymbol\theta})=-2\sum_{i=1}^m I(X_t=i)\log (\hat{\chi}_t)_i=-2\log(\hat{\chi}_t)_{X_t}$, also called cross-entropy. We remind that $\hat\chi_t$ is the $m$-probability vector describing the prediction of the model and in the previous formula we take the $X_t$-th component. For the MTDg(p) model this probability vector  is
\begin{equation}
\hat{\chi}_t=\sum_{g=1}^p\chi_{t-g}  \lambda_g\boldsymbol{Q}^g
\end{equation}
where, as before, $\chi_{t-g}$ is a $m$-vector of zeros with the exception of the realized component $X_{t-g}$. This quantity can be easily computed once the model is calibrated, since it depends on the transition probabilities. EPE values are in the range $[0,+\infty)$, and it is zero if all probabilities $(\hat{\chi}_t)_{X_t}$ of the sample are equal to one (perfect prediction), and it is infinity if all probabilities $(\hat{\chi}_t)_{X_t}$ of the sample are zero (prediction of impossible events). 

We evaluate the best performing model as the model with the lowest EPE and benchmark the MTDg  with a model with the unconditional probabilities as predictors of future signed events.  Table~\ref{tab:epe} reports all the EPE values for different models of the predictor estimated on MSFT, Bank of America-CitiGroup (BAC), General Electric (GE), Cisco (CSCO), AAPL, and Amazon (AMZN) data. The scheme of the out-of-sample analysis is the following: The model is trained on a time period of 10 days, then we compute the loss functions in the following trading day by using the parameter set provided by MLE (strongly constrained) or by GMM (weakly constrained). We repeat the procedure by shifting the estimation by one trading day ahead. Finally, we compute the global loss by averaging all measured loss function. The financial interpretation of the EPE values is clear in the case of the large tick stocks, because a price-changing event moves the price by one tick with probability almost one and thus there exists a direct relation between the states of the MTDg model and the price return. Hence, for large tick stocks the EPE value can be employed as a proxy of the predictability of returns at high frequency time scale.
  \begin{table}
    \centering
    \begin{tabular}{cc|c|c|c}
      \hline
      \hline
      & & Model A & Model B & Model C \\
      \hline \hline
      \multirow{2}{*}{MSFT} & EPE & 1.928 & 1.199 & 1.181 \\ 
      & SE & 0.003 & 0.004 & 0.004 \\
      \hline 
      \multirow{2}{*}{BAC} & EPE & 1.744 & 0.799 & 0.785 \\ 
      & SE & 0.003 & 0.004 & 0.004 \\ 
      \hline 
      \multirow{2}{*}{GE} & EPE & 1.922 & 1.169 & 1.153 \\ 
      & SE & 0.004 & 0.005 & 0.005 \\
      \hline
      \multirow{2}{*}{CSCO} & EPE & 1.919 & 1.112 & 1.098 \\ 
      & SE & 0.004 & 0.005 & 0.005 \\
      \hline
      \multirow{2}{*}{AAPL} & EPE & 2.643 & 2.211 & 2.192 \\ 
      & SE & 0.001 & 0.002 & 0.002 \\
      \hline
      \multirow{2}{*}{AMZN} & EPE & 2.579 & 2.196 & 2.183 \\ 
      & SE & 0.002 & 0.004 & 0.004 \\
      \hline \hline
    \end{tabular}
    \caption{EPE values and standard errors (SE) for MSFT, BAC, GE, CSCO, AAPL and AMZN data. \textit{Model A}: Unconditional probabilities as predictor. \textit{Model B}: Strongly constrained MTDg(100) estimated via MLE according to Eq.~\ref{eqn:mtd_exp}. Total number of parameters: 11. \textit{Model C}: Weakly constrained MTDg(100) model estimated via GMM with matrices as in Eq.~\ref{eqn:smart_mtdg_model}. Total number of parameters: $500$.}
    \label{tab:epe}
  \end{table}

From Table~\ref{tab:epe} we see that both MTDg models out-perform the benchmark. More importantly, there is a clear evidence that the weakly constrained model with the highest number of parameters (Model C) outperforms the strongly constrained MTDg, for all considered stocks. These results exclude the over-fitting hypothesis, and support the claim that weakly constrained MTDg models are good candidates to capture the high-frequency dynamics of signed events.

\section{Discussion and conclusion}
\label{sec:conclusions}
The companion paper~\cite{taranto16} has established that treating all market orders on the same basis produces erroneous predictions both for the ``response functions'' (average lagged impact)  at negative lags and the signature plot. Single-propagator models and history dependent impact models are not designed to capture the feedback effects between past price returns and future order flow. These serious discrepancies have been significantly reduced by introducing the extended versions of the linear impact models (TIM and HDIM) which consider a richer set of signed events (see~\cite{eisler2012a, eisler2012b}). The argument which has motivated our generalization of the impact models is the observation that price-changing and non price-changing events have to be treated differently. This is particularly evident for large tick stocks, where price moving events are extremely rare but very informative. This apparently minor modification has lead to an extended class of propagator models which describe with remarkable realism the intertwined high-frequency dynamics of prices and order flow. Nonetheless, the linear description of the market dynamics achieved in Part I~\cite{taranto16} is still too rigid: these models are designed to describe the evolution of the market with an exogeneously specified order flow. This fact seriously limits the forecasting capabilities of linear impact models. 

The Mixture Transition Distribution model partly solves the above issue by introducing an explicit stochastic model for the order flow, treated as an endogenous component of the dynamics. It is specially designed for variables which are inherently discrete -- a feature of great relevance for price returns of large tick stocks. In this paper we have presented a class of so-called MTDg models as a natural extension of the Discrete Autoregressive DAR(p) in a multi-event context. Our aim was to test how well a calibrated MTDg model can account for the statistics of the order flow, i.e. the string of 4 events: buy/sell -- price changing/non changing events. One of the most interesting aspects of our work is of methodological nature, and concerns the practical calibration of ``large'' models. The class of weakly constrained MTDg models introduced in section~\ref{sec:weakly} and Appendix~\ref{app:C} represents a rich family of discrete models, where the number of free parameters equates the number of independent observable correlation functions. This fact allows to introduce a numerical procedure which solves the estimation of the model parameters in a remarkably robust way. This result is rooted on the proof that the optimization problem is convex in the parameter space. From the financial viewpoint we have shown that -- perhaps surprisingly -- a weakly constrained version of the MTDg models captures the dynamics of signed events with greater realism than alternative and more parsimonious versions. Despite the large number of parameters, the out-of-sample analysis confirms that such good performances are achieved without over-fitting the data.

The improvement brought by the MTDg models and the new estimation methodology is remarkable, but still some discrepancies persist when comparing the model predictions with the empirical correlation functions. Several reasons may be responsible for these deviations. The first one was already pointed out by Raftery in~\cite{raftery1985} where he has shown that there exist regions of correlations which simply cannot be reproduced by MTDg models. A second reason is that, even though the MTDg model was the correct data generating process, the estimation methodology which involves information only coming from second order conditions, may lack efficiency with respect to the MLE approach. Finally, the MTDg model represents a parsimonious approximation of a full Markov chain of order $p$. This parsimony may come at expense of the realism of the model. 

From a microstructural point of view, we can hypothesize that the string of past signed events $X_{t-1},\ldots,X_{t-p}$ is not informative enough to predict the value $X_t$. In particular, for large tick stocks price-changing events $\pi=\mathrm{C}$ are much rarer than non price-changing event $\pi=\mathrm{NC}$. Therefore, a $\pi=\mathrm{C}$ event is by construction difficult to predict with past information based only on realised signs and trades. Hence, the behavior that we observe may be ascribed to a problem of missing explanatory variables. A natural candidate in this respect could be the volume of orders outstanding at the opposite side of the limit order book before the execution of a trade order, i.e. the local order book imbalance.

From a more fundamental point of view, we should also point out that the MTDg calibrated kernel, which gives the probability that an event at time $t=0$ will trigger similar or opposite events at time $t=g$ later, must be interpreted with care. Indeed, this kernel receives contributions both from order splitting, which increases the probability that an agent places an order of the same sign in the future, and from genuine reactions of the rest of the market to this event~\cite{toth2015,toth2012}. These reactions can be herding (copy cat trades) or, on the contrary, trades in the other direction (coming e.g. from liquidity providers). The response of the order flow to a single, isolated trade is thus expected to be rather different from the impulse function obtained by calibrating an MTDg model to the full order flow since order splitting contributions will be absent in the former, but contribute to the latter. The distinction between the two effects requires trade identification to be resolved. We hope to come back to this issue in a forthcoming work~\cite{toth2016}. 

\section*{Acknowledgement}
We want to thank Z. Eisler, J. Donier, and I. Mastromatteo for very useful discussions. D. E. Taranto acknowledges CFM for supporting his extended visit at CFM where part of this research was done.

\appendix
\section{Existence and uniqueness of the stationary distribution of the MTDg model}
\label{app:B}
\begin{theorem}
Suppose that a sequence of random variables $\left\lbrace X_t\right\rbrace_{t\in \mathbb{N}}$ taking values in the finite set $\mathcal{X}=\lbrace 1,\ldots,m\rbrace$ is defined by 
\begin{align*}
\mathbb{P}(X_t=i|X_{t-1}=i_1,\ldots,X_{t-p}=i_p) = \sum_{g=1}^p \lambda_g q^g_{i_g,i}\,,
\end{align*}
where $\boldsymbol{Q}^g=\left[q_{i,j}^g\right]_{i,j\in \mathcal{X}}$ are matrices with normalized row, $\sum_j q^g_{i,j}=1$,
$\sum_{g=1}^p \lambda_g=1$, and assume that $\hat{\eta} \boldsymbol{Q}^g=\hat{\eta}, \forall g$. If the vector $\hat{\eta}$ is such that $\hat{\eta}_i>0, i \in \mathcal{X}$ and $\sum_i \hat{\eta}_i=1$, and
\begin{align}
0 < \sum_{g=1}^p \lambda_g q_{i_g,i}^g < 1, \qquad \forall (i,i_1,\ldots,i_p)\in \mathcal{X}^{p+1}\,,
\label{eqn:appB_positivity}
\end{align}
then
\begin{align*}
\lim_{\ell \to \infty} \mathbb{P}(X_{t+\ell}=i|X_{t-1}=i_1,\ldots,X_{t-p}=i_p)=\hat{\eta}_{i}\,.
\end{align*}
\end{theorem}
\textbf{Proof.}
Let $\boldsymbol{T}$ be the $m^p \times m^p$ transition matrix for the Markov chain with the $m^p$ possible values of $(X_{t-1},\ldots,X_{t-p})$ as states. The elements of $\boldsymbol{T}$ are
\begin{align*}
\mathbb{P}(X_t=i,X_{t-1}=i_1,\ldots,X_{t-p+1}=i_{p-1}|X_{t-1}=j_1,\ldots,X_{t-p}=j_p)\,, \\
=\begin{cases}
\sum_{g=1}^p \lambda_g q_{j_g,i}^g & \mbox{if } i_g=j_g, \mbox{ for } g=1,2,\ldots,p-1\,, \\
0 & \mbox{otherwise}\,.
\end{cases}
\end{align*}
Each column of $\boldsymbol{T}$ represents the $p$-vector $(i,\ldots,i_{p-1})$ of arrival states, which are ordered in such a way that $i$ varies most slowly, $i_1$ second most slowly, and so on. Similarly, the rows of $\boldsymbol{T}$ represents the values of $(j_1,\ldots,j_p)$ with $j_1$ varies most slowly, and so on.

The assumption of Eq.~\ref{eqn:appB_positivity} guarantees that all states of $\boldsymbol{T}$ intercommunicate, so $\boldsymbol{T}$ is irreducible. Amongst the diagonal elements of $\boldsymbol{T}$ $m$ are aperiodic, then, since $\boldsymbol{T}$ is irreducible, all states are aperiodic. Hence, $\boldsymbol{T}$, being finite, specifies an ergodic Markov chain and has a unique equilibrium distribution $\xi$ satisfying $\xi \boldsymbol{T}=\xi$ with elements
\begin{align*}
\xi_{j_1,\ldots,j_p}=\lim_{t \to \infty} \mathbb{P}(X_{t-1}=j_1,\ldots,X_{t-p}=j_p)
\end{align*}
where the $p$-vector $(j_1,\ldots,j_p)$ is ordered in the same way of the matrix $\boldsymbol{T}$. We call $\omega=(\omega_1,\ldots,\omega_m)$ the corresponding one-dimensional marginal equilibrium distribution. Also let $\boldsymbol{R}$ be the ``collapsed form'' of $\boldsymbol{T}$ as defined in~\cite{pegram1980}, which is the $m^p \times m$ matrix of the non-zero elements of $\boldsymbol{T}$. Clearly, in general
\begin{align}
\xi \boldsymbol{R} = \omega\,.
\label{eqn:appB_existence}
\end{align}
We write the same matrix for the model (\ref{eqn:smart_mtdg})
\begin{align*}
\boldsymbol{R} = \sum_{g=1}^p \lambda_g \boldsymbol{U}_g\,,
\end{align*}
where $\boldsymbol{U}_g=\boldsymbol{A}_{g,1} \otimes \cdots \otimes \boldsymbol{A}_{g,p}$ where
\begin{align*}
\boldsymbol{A}_{g,k}=\begin{cases}
\boldsymbol{Q}^g & \mbox{if } g=k\\
1^T & \mbox{if } g \neq k
\end{cases}
\end{align*}
and $\otimes$ is the Kronecker product and $1^T$ is a $m \times 1$ vector of ones.

We now calculate $\xi \boldsymbol{R}$ in another way. The $k$-th column of $\xi \boldsymbol{U}_g$ is
\begin{align*}
\sum_{i_1,\ldots,i_p}^m  q_{i_g,k}^g \xi_{i_1,\ldots,i_p}&=\sum_{i_g}^m q_{i_g,k}^g \sum_{i_h, h \neq g}^m \xi_{i_1,\ldots,i_p} \\
&=\sum_{i_g}^m q_{i_g,k}^g \omega_{i_g}
\end{align*}
which is also the $k$-th column of $\omega \boldsymbol{Q}^g$. Thus
\begin{align}
\xi \boldsymbol{R}=\sum_{g=1}^p \lambda_g \omega \boldsymbol{Q}^g\,.
\label{eqn:appB_uniqueness}
\end{align}
Equating Eqs.~\ref{eqn:appB_existence} and \ref{eqn:appB_uniqueness}, we have that $\omega=\hat{\eta}$, by uniqueness of $\omega$ and $\hat{\eta} \boldsymbol{Q}^g=\hat{\eta}, \forall g$.

\section{System of matrix equations of the MTDg model}
\label{app:A}
\begin{proposition}
  Suppose that a sequence of random variables $\left\lbrace X_t\right\rbrace_{t\in \mathbb{N}}$ taking values in the finite set $\mathcal{X}=\lbrace 1,\ldots,m\rbrace$ is defined by Eq.~\ref{eqn:mtdg} and is stationary. Let $\boldsymbol{B}(k)$ be a $m \times m$ matrix with elements
  \begin{align*}
    b_{i,j}^k=\mathbb{P}(X_t=i, X_{t+k}=j), \qquad i,j \in \mathcal{X}; k\in \mathbb{Z}
  \end{align*}
  and $\boldsymbol{B}(0)=\mbox{diag}(\hat{\eta}_1,\ldots,\hat{\eta}_m)$. Then
  \begin{align*}
    \boldsymbol{B}(k)=\sum_{g=1}^p \lambda_g \boldsymbol{B}(k-g) \boldsymbol{Q}^g\,.
  \end{align*}
\end{proposition}
\textbf{Proof.}
First consider the case where $k=1,\ldots,p$. Let
\begin{align*}
Y_t^k&=\lbrace X_{t+k-g}: g=1,\ldots,p; g \neq k\rbrace,
\end{align*}
then
\begin{align*}
b_{i,j}^k&=\mathbb{P}(X_t=i, X_{t+k}=j) \\
&=\sum_{Y_t^k}\mathbb{P}(X_t=i, X_{t+k}=j|Y_t^k)\mathbb{P}(Y_t^k) \\
&=\sum_{Y_t^k}\mathbb{P}(X_{t+k}=j|X_t=i,Y_t^k)\mathbb{P}(X_t=i|Y_t^k)\mathbb{P}(Y_t^k) \\
&=\sum_{Y_t^k} \sum_{g=1,g \neq k}^p \lambda_g q_{X_{t+k-g},j}^g \mathbb{P}(X_t=i|Y_t^k)\mathbb{P}(Y_t^k)+\sum_{Y_t^k}\lambda_k q_{i,j}^k \mathbb{P}(X_t=i|Y_t^k)\mathbb{P}(Y_t^k) \\
&=\sum_{g=1,g \neq k}^p \lambda_g \sum_{X_{t+k-g}} q_{X_{t+k-g},j}^g \mathbb{P}(X_t=i|X_{t+k-g})\mathbb{P}(X_{t+k-g}) + \lambda_k \hat{\eta}_i q_{i,j}^k \\
&=\sum_{g=1,g \neq k}^p \lambda_g  \sum_{h=1}^m b_{i,h}^{k-g} q_{h,j}^g + \lambda_k \hat{\eta}_i q_{i,j}^k
\end{align*}
which is the $(i,j)$-th element of
\begin{align*}
\sum_{g=1}^p \lambda_g \boldsymbol{B}(k-g) \boldsymbol{Q}^g
\end{align*}
as required.

\section{A general class of MTDg models}
\label{app:C}
Let $\boldsymbol{B}(k)$ be an $m \times m$ matrix whose elements are
\begin{equation*}
b_{i,j}^k=\mathbb{P}(X_t=i, X_{t+k}=j), \qquad i,j=1,\ldots,m, k \in \mathbb{Z},
\end{equation*}
where $\boldsymbol{B}(0)=diag(\hat{\eta}_1,\ldots,\hat{\eta}_m)$. The matrices $\boldsymbol{B}(k)$ represent the bivariate distributions of the random variable $X_t$. Then, we have that
\begin{align*}
\boldsymbol{B}(k)=\begin{pmatrix}
b_{1,1}^k & \cdots & b_{1,m-1}^k & \hat{\eta}_1-\sum_{i=1}^{m-1}b_{1,i}^k \\
\vdots & \ddots & \vdots & \vdots \\
b_{m-1,1}^k & \cdots & b_{m-1,m-1}^k & \hat{\eta}_{m-1}-\sum_{i=1}^{m-1}b_{m-1,i}^k \\
\hat{\eta}_1-\sum_{i=1}^{m-1}b_{i,1}^k & \cdots & \hat{\eta}_{m-1}-\sum_{i=1}^{m-1}b_{i,m-1}^k & 2\hat{\eta}_m-1+\sum_{i,j=1}^{m-1} b_{i,j}^k \\
\end{pmatrix}\,, 
\end{align*}
where the total number of independent elements is $m^2-2m+1$ for each $k$. The parameters of the MTD model of order $p$ consists in the vector $\lambda=(\lambda_1,\ldots,\lambda_p)$ and the matrices $\boldsymbol{Q}^g$, such that
\begin{align*}
\boldsymbol{Q}^g&=1^T\hat{\eta}+\boldsymbol{\tilde{Q}}^g\,, \nonumber \\
\boldsymbol{\tilde{Q}}^g&=\begin{pmatrix}
\tilde{q}_{1,1}^g & \cdots & \tilde{q}_{1,m-1}^g & -\sum_{i=1}^{m-1}\tilde{q}_{1,i}^g \\
\vdots & \ddots & \vdots & \vdots \\
\tilde{q}_{m-1,1}^g & \cdots & \tilde{q}_{m-1,m-1}^g & -\sum_{i=1}^{m-1}\tilde{q}_{m-1,i}^g \\
-\sum_{i=1}^{m-1}c_i \tilde{q}_{i,1}^g & \cdots & -\sum_{i=1}^{m-1}c_i \tilde{q}_{i,m-1}^g & \sum_{i,j=1}^{m-1} c_i \tilde{q}_{i,j}^g\\
\end{pmatrix}\,, 
\end{align*}
where $\hat{\eta}\boldsymbol{\tilde{Q}}^g=0, \forall g$ and $c_i=\hat{\eta}_i/\hat{\eta}_m$. Consistently with these definitions, the conditional probabilities of the $p$-order Markov chain read
\begin{align*}
\mathbb{P}(X_t=i|X_{t-1}=i_1,\ldots,X_{t-p}=i_p) = \hat{\eta}_{i}+\sum_{g=1}^p a_{i_g,i}^g\,,
\end{align*}
where $a_{i_g,i}^g \equiv \lambda_g \tilde{q}_{i_g,i}^g$.

Within this framework, the bivariate distributions and the matrices $\boldsymbol{\tilde{Q}}^g$ satisfy the following system of matrix equations
\begin{align*}
\boldsymbol{B}(k)-\hat{\eta}^T\hat{\eta}=\sum_{g=1}^p \boldsymbol{B}(k-g) \boldsymbol{A}_g\,,
\end{align*}
where $\boldsymbol{A}^g \equiv \lambda_g\boldsymbol{\tilde{Q}}^g$. Employing the empirical bivariate distributions, above linear system can be inverted in order to find the parameters of the model. Resulting parameters have to satisfy the following conditions in order to characterise a well defined $p$-order Markov model 
\begin{align*}
  \hat{\eta}_{i}+\sum_{g=1}^p \max_{i_g} \left(a_{i_g,i}^g \right) < 1, \qquad \forall i \in \mathcal{X}\,;  \nonumber \\
  \hat{\eta}_{i}+\sum_{g=1}^p \min_{i_g} \left(a_{i_g,i}^g \right) > 0, \qquad \forall i \in \mathcal{X}\,.
\end{align*}

\section{Convexity of the optimization problem}
\label{app:D}

\begin{proposition}
If $\boldsymbol{K}$ is not singular, the following constrained optimization problem
\begin{align*}
  \hat{q}&=\underset{{\bf q} \in \mathbb{R}^{p(m^2-2m+1)}}{\operatorname{argmin}} \left\Vert {\bf d}-\boldsymbol{K}\cdot {\bf q} \right\Vert^2 \nonumber \\
  \mbox{s.t.} \qquad &\hat{\eta}_{i}+\sum_{g=1}^p \max_{i_g} \left(a_{i_g,i}^g \right) < 1, \qquad \forall i \in \mathcal{X}  \nonumber \\
  &\hat{\eta}_{i}+\sum_{g=1}^p \min_{i_g} \left(a_{i_g,i}^g \right) > 0, \qquad \forall i \in \mathcal{X} 
\end{align*}
is convex in $\mathbb{R}^{p(m^2-2m+1)}$. 
\end{proposition}

\textbf{Proof.} This is true if the objective function and all the constraints are convex functions. First of all, it is straightforward to show that the Hessian of the objective function $2\boldsymbol{K}\boldsymbol{K}^T$ is a positive semi-definite matrix. The constraints are convex in $q$, if they are convex in the parameters $a_{i,j}^g$ because they are affine functions of the components of $q$. Let $a$ be the vector of parameters $\left(a_{i,j}^g \right)_{i,j \in \mathcal{X}; 1 \leq g \leq p}$, we need to prove that the function
\begin{align*}
  f(a)=\sum_{g=1}^p \max_{i_g} \left(a_{i_g,i}^g \right), \qquad \forall i \in \mathcal{X}
\end{align*}
is convex in $\mathbb{R}^{p(m^2-2m+1)}$.

If we prove it for a fixed $i$, then it is true for all $i \in \mathcal{X}$ and also for the constraints with the minimum function. The function $f(a)$ satisfies, for $0 \leq \theta \leq 1 $, different vectors of parameters $a,b\in \mathbb{R}^{m^2p}$, and fixed $i$
\begin{align*}
  f(\theta a + (1-\theta)b)&=\sum_{g=1}^p \max_{i_g} \left(\theta a_{i_g,i}^g+(1-\theta) b_{i_g,i}^g \right) \nonumber \\
  &\leq\theta \sum_{g=1}^p \max_{i_g} \left(a_{i_g,i}^g \right) + (1-\theta) \sum_{g=1}^p \max_{i_g} \left(b_{i_g,i}^g \right) \nonumber \\
  &=\theta f(a)+(1-\theta)f(b)\,.
\end{align*}
Therefore, we conclude that the function $f(a)$ is convex in $\mathbb{R}^{p(m^2-2m+1)}$.

\end{document}